\documentclass{article}

\usepackage{amssymb,amsfonts,amsmath}
\usepackage{cite,enumerate,float}
\usepackage{color}
\usepackage{tikz}
\usetikzlibrary{arrows,snakes,backgrounds}
\usepackage{url}

\def\be{\begin{eqnarray}}
\def\ee{\end{eqnarray}}
\def\nn{\nonumber}

\def\p{\partial}

\def\Tr{{\rm Tr}\,}
\def\wc{weak composition}
\def\wcs{weak compositions}

\definecolor{red}{rgb}{1,0,0}
\definecolor{orange}{rgb}{1,0.5,0}
\definecolor{violet}{rgb}{0.7,0,1}

\newcommand\br[1]{
  \left(#1\right)
}
\newcommand\mfC[0]{\mathfrak{C}}
\newcommand\ux[0]{\underline{x}}

\hoffset= - 1.0in         

\begin{document}

\title{\vspace{1.5cm}\bf
Twisted Cherednik systems and non-symmetric Macdonald polynomials
}

\author{
A. Mironov$^{b,c,d,}$\footnote{mironov@lpi.ru,mironov@itep.ru},
A. Morozov$^{a,c,d,}$\footnote{morozov@itep.ru},
A. Popolitov$^{a,c,d,}$\footnote{popolit@gmail.com}
}

\date{ }

\maketitle

\vspace{-6cm}

\begin{center}
  \hfill MIPT/TH-20/25\\
  \hfill FIAN/TD-21/25\\
  \hfill ITEP/TH-28/25\\
  \hfill IITP/TH-25/25
\end{center}

\vspace{4.5cm}

\begin{center}
$^a$ {\small {\it MIPT, Dolgoprudny, 141701, Russia}}\\
$^b$ {\small {\it Lebedev Physics Institute, Moscow 119991, Russia}}\\
$^c$ {\small {\it NRC ``Kurchatov Institute", 123182, Moscow, Russia}}\\
$^d$ {\small {\it Institute for Information Transmission Problems, Moscow 127994, Russia}}
\end{center}

\vspace{.1cm}

\begin{abstract}
We consider eigenfunctions of many-body system Hamiltonians associated with generalized ($a$-twisted) Cherednik operators used in construction of other Hamiltonians: those arising from commutative subalgebras of the Ding-Iohara-Miki (DIM) algebra. The simplest example of these eigenfunctions is provided by {\it non-symmetric} Macdonald polynomials, while generally they are constructed basing on the ground state eigenfunction coinciding with the twisted Baker-Akhiezer function being a peculiar (symmetric) eigenfunction of the DIM Hamiltonians. Moreover, the eigenfunctions admit an expansion with universal coefficients so that the dependence on the twist $a$ is hidden only in these ground state eigenfunctions, and we suggest a general formula that allows one to construct these eigenfunctions from non-symmetric Macdonald polynomials.
This gives a new twist in theory of integrable systems,
which usually puts an accent on {\it symmetric} polynomials, and
provides a new dimension to the {\it triad} made from the symmetric Macdonald polynomials, untwisted Baker-Akhiezer functions
and Noumi-Shiraishi series.
\end{abstract}

\bigskip

\newcommand\smallpar[1]{
  \noindent $\bullet$ \textbf{#1}
}

\section{Introduction}

Typical many-body integrable systems are systems of Calogero-Moser-Satherland and Ruijsenaars-Schneider families, and they
have Schur-Jack-Macdonald symmetric polynomials as their typical eigenfunctions \cite{Jack,St,Turb,Mac}.
Since nowadays the hidden integrability is understood to be a guiding principle for
description of non-perturbative functional integrals and $D$-modules
associated to them \cite{UFN2,UFN3}, the deep algebraic structure behind Macdonald theory
is attracting more and more attention in mathematical physics.
The underlying symmetry here is the Ding-Iohara-Miki (DIM) algebra \cite{DI,Miki}, or equivalently, the elliptic Hall algebra \cite{K,BS,S} (which is basically the same \cite{S,Feigin}), which actually involves many more integrable systems than Calogero-Moser-Satherland and Ruijsenaars-Schneider families \cite{MMP}, each being associated with a ray passing through any integer point $(kn,km)$ on the $2d$ integer plane ($n$ and $m$ are coprime). This justifies a growing attention to DIM representation theory.

$N$-body representations of DIM are controlled by some Cherednik type operators $Ch_i^{(n,m)}$, $i=1,\ldots,N$ commuting at fixed $n$ and $m$ but distinct $i$, so that all the commuting Hamiltonians of integrable systems associated with the ray $(n,m)$, $H_k^{(n,m)}$ are manifestly constructed \cite{MMP} as symmetrization of the sums $\sum_{i=1}^N \Big(Ch_i^{(n,m)}\Big)^k$, which, to some extent, reminds conventional Casimir operators ${\rm Cas}_k = \Tr T^k$:
\be\label{1}
H_k^{(n,m)}=\hbox{Sym}\left(\sum_{i=1}^N \Big(Ch_i^{(n,m)}\Big)^k\right)
\ee

Symmetric Macdonald polynomials are just particular eigenfunctions of the Hamiltonians $H_k^{(0,1)}$,
while their general solutions are described by a whole {\it triad} \cite{triad},
involving also non-symmetric Baker-Akhiezer functions \cite{Cha,ChE,MMP1} and
Noumi-Shiraishi series \cite{NS}.

In this paper, we propose to study not only the Hamiltonians $H_k^{(n,m)}$ associated with ray $(n,m)$ of the DIM algebra, but also another integrable system associated with the same ray: that with Hamiltonians $Ch_i^{(n,m)}$. In the case of ray $(0,1)$, the eigenfunctions of these Cherednik operators are just non-symmetric Macdonald polynomials.

The two integrable systems express a relation \cite{DIMDAHA} between the Elliptic Hall (or DIM) and spherical DAHA \cite{Ch} algebras, and, because of (\ref{1}), the eigenfunctions of Hamiltonians of these two systems coincide when they are symmetric.

Now note that not symmetric but quasipolynomial eigenfunction of Hamiltonians $H_k^{(n,m)}$ is known \cite{MMP1} to be twisted Baker-Akhiezer function \cite{ChE,ChF}. Because of this, we call the system of commuting $Ch_i^{(n,m)}$ twisted Cherednik integrable system. If one now looks for a ``ground state'' solution of this Cherednik system, it is symmetric as any ground state and, hence, it has simultaneously be an eigenfunction of the $H_k^{(n,m)}$ Hamiltonians because of (\ref{1}), i.e. it has to be simultaneously a particular Baker-Akhiezer function that is symmetric! We will demonstrate that it is really the case.

We begin from an elementary example in sec.\ref{toy}, where an order in the world of polynomials is introduced
by interpreting them as eigenvalues of simplest differential operators.
The main point here is coexistence of two kind of operators like generators of $U(1)^N$ and
the would be ``Casimir operators'' made from ``traces'' of their powers.
Then, in sec.\ref{nspol}, we briefly describe the construction of non-symmetric polynomials as a kind od Verma module build by
action of creation operators, not obligatory differential.
This is the method used in one half of existing literature, and is the main approach to the key, Demazure and Schubert families.
After that, we return to the eigenfunction approach, promoting simple dilatations to more sophisticated
Cherednik operators. This approach leads to non-symmetric Macdonald polynomials in sec.\ref{nsMac}, though
the key and Demazure polynomials are also naturally embedded into this scheme, used in another half of the current literature,
while Schubert requires additional considerations within this approach.

Then, in sec.\ref{nsMac}, we discuss integrable systems associated with DIM algebra, their Hamiltonians and corresponding eigenfunctions. These eigenfunctions are twisted Baker-Akhiezer functions, some of them also emerging as particular (ground state) eigenfunctions of Hamiltonians of the twisted Cherednik integrable systems discussed in sec.\ref{intCh}. A detailed description of the eigenfunctions of the twisted Cherednik Hamiltonians is contained in sec.\ref{efCH}. The last section contains a summary and discussion, and, in the Appendix, we study the limit of $q,t\to 1$ holding $\beta:=\log t/\log q$ fixed. In this limit, the twisted system is obtained from the non-twisted one (where the eigenfunctions are just non-symmetric Jack polynomials) just with multiplication (twisting) by a simple function, however, some formulas remain rather instructive (and easier to deal with) even in this trivial limit.

Last but not least: we construct the eigenfunctions of the twisted Cherednik Hamiltonians in secs.\ref{intCh},\ref{efCH}, at $t=q^{-m}$, $m\in\mathbb{Z}_{\ge 0}$. We sometimes write down possible continuation to arbitrary $t$, however, it is ambiguous (see sec.\ref{5.2}).

\paragraph{Notation.}
The $q$-Pochhammer symbols are standardly defined
\be\label{Poch}
(x;q)_\infty:&=&\prod_{j=0}^\infty(1-q^jx)\nn\\
(x;q)_n:&=&{(x;q)_\infty\over(xq^n;q)_\infty}=\prod_{j=0}^{n-1}(1-q^jx)=\sum_{k=0}^n(-1)^kq^{k(k-1)\over 2}\binom{n}{k}_qx^k
\ee
and $\binom{n}{k}_q$ are $q$-binomial coefficients.

The integer part of a number $x$ is denoted through $[x]$, while the $q$-number is
\be
[x]_q:={q^x-1\over q-1}
\ee

Throughout the paper, if $\lambda$ is the weak integer composition, i.e. a vector with non-negative components $\{\lambda_i\}$, we denote through $\lambda^+$ the corresponding ordered partition, i.e. the vector with ordered components $\lambda_1\ge\ldots\lambda_n\ge 0$. If one associates $\lambda$ with a point of the integral weight lattice of $GL_n$, $\lambda^+$ corresponds to the associated dominant integral weight.

\section{Warm-up examples}

\subsection{A toy example
\label{toy}}

We start with a toy example, which demonstrates our main ideas in this paper in full.

Consider the system of commuting operators
\be
\hat c_i = x_i\frac{\p}{\p x_i},\ \ \ \  i = 1,\ldots, N
\ee
Their common eigenfunctions are
\be
e_\lambda(x) = x^\alpha:= \prod_{i=1}^N x_i^{\lambda_i}
\ee
labeled by arbitrary sets of $N$ non-negative integers (weak compositions).

Among these common eigenfunctions there are no symmetric polynomials.
Symmetric polynomials are, however, among eigenfunctions of another commutative system: that of operators
\be
\hat h_k = \sum_{i=1}^n \hat c_i^k
\ee
Symmetric eigenfunctions are labeled by the Young diagrams (ordered partitions) $\lambda^+$,
i.e. {\it ordered} sequences of positive integers $\lambda_1\geq\lambda_2\geq \ldots >0$,
describing \wcs of the {\it level} $|\lambda|:=\sum_i\lambda_i$:
\be
s_{\lambda^+}(x) = \prod_{a=1} \left(\sum_i x_i^{\lambda_a}\right) = p^\lambda:=\prod_a p_{\lambda_a}
\ee
where the {\it time-variables, restricted to Miwa locus}, are $p_k[x]:=\sum_{i=1}^N x_i^k$.

\bigskip

Both non-symmetric and symmetric polynomials satisfy Cauchy identities:
\be
\sum_\lambda e_\lambda(x)e_\lambda(y) = \prod_{i=1}^N \left(\sum_{\lambda_i=0}^\infty (x_iy_i)^{\lambda_i}\right)
= \prod_{i=1}^N \frac{1}{1-x_iy_i}
\label{eCauchy}
\ee
and
\be
\sum_{\lambda^+} {s_{\lambda^+}(x)s_{\lambda^+}(y)\over z_{\lambda^+}} =
\prod_{i,j}\frac{1}{1-x_iy_j} = \exp\left(\sum_{k=1}^\infty\frac{p_k[x] p_k[y]}{k}\right)
\ee
where $z_{\lambda^+}:=\prod_k k^{m_k}m_k!$ is the order of automorphism of the Young diagram $\lambda^+$, and $m_k$ is the number of lines of length $k$ in the Young diagram $\lambda^+$.

The shape of Cauchy identity depends on normalization of polynomials, for example,
(\ref{eCauchy}) can be changed for
\be
\sum_\lambda \frac{e_\lambda(x)e_\lambda(y)}{\prod_i \lambda_i!}
\prod_{i=1}^N \left(\sum_{\lambda_i=0}^\infty \frac{(x_iy_i)^{\lambda_i}}{\lambda_i!}\right)
= \exp\left(\sum_{i=1}^N x_i y_i\right)
\ee
or
\be
\sum_\lambda \frac{e_\lambda(x)e_\lambda(y)}{|\lambda|} =  \int_0^\infty dz  \prod_{i=1}^N \left(\sum_{\lambda_i=0}^\infty (x_iy_ie^{-z})^{\lambda_i}\right) =\int_0^\infty dz \ \prod_{i=1}^N \frac{1}{1-x_iy_ie^{-z}}
\ee

\bigskip

Symmetric polynomials and their Hamiltonians $\hat h_k$ have an interesting set of deformations
to Schur-Jack-Macdonald polynomials, which are eigenfunctions of Calogero-Ruijsenaars Hamiltonians $\hat H_k$.
Analogously, deformations exist for non-symmetric polynomials,
now named\footnote{A separate story is about the Schubert polynomials.} {\it key}, {\it Demazure} and {\it non-symmetric Macdonald},
and, for operators $\hat c_i$, they become Cherednik operators $\hat C_i$.

However, the logic and even the details of the construction remain literally the same as they were for the toy example of $\hat c_i$.
One may say that we just switch to another basis, satisfying another kind of orthogonalization conditions,
which is related by a conjugation with a kind of Vandermonde determinant and its $q,t$-deformations.
Still the theory becomes/looks pretty sophisticated.

\subsection{Constructing non-symmetric polynomials\label{nspol}}

Construction of non-symmetric polynomials generalizing $e_\lambda$ can be done in a few natural ways. One way, which we briefly describe in this subsection, is to use an iterative construction. They can be also constructed from orthogonality relations, and as eigenfunctions of a commutative set of operators. We discuss these ways in sec.\ref{nsMac} in the example of non-symmetric Macdonald polynomials.

\subsubsection{Iterative construction of non-symmetric polynomials}

The non-symmetric polynomial depends on $n$ variables $x_1,\ldots,x_n$ and on the permutation $w$ from permutation group ${\cal S}_n$:
\be
P_{\lambda^+,w}[x] = {\cal P}_\lambda[x]
\ee
Here $\lambda^+$ is a Young diagram with $\lambda_1\geq \lambda_2 \geq \ldots$
and $\lambda = \hat w\circ \lambda^+$ is a {\it disordered} sequence made from the same $\lambda_a$
(called weak composition).
Most families of non-symmetric polynomials are generated by operators $\hat \pi_i$,
\be
P_{\lambda^+,w}[x] = \hat \pi_w  x^{\lambda^+}
\ee
with $\hat w = \prod_I \sigma_{I}$,
$\hat \pi_w := \prod_I \hat\pi_I$
and $x^\lambda = \prod_{i=1}^n x_i^{\lambda_i}$.
Here $\sigma_i:=\hat \sigma_{i,i+1}$ is the permutation of two adjacent variables $x_i$ and $x_{i+1}$,
and the same $\sigma_i$ may appear in $\hat w$ many times, hence we denote it by a different letter $I$.
Sometimes, $\lambda^+$ is fixed to be $\lambda_0:=[n-1,n-2,\ldots,2,1]$ so that $x^{\lambda_0}=x_1^{n-1}x_2^{n-2} \ldots x_{n-1}$.

\bigskip

Various operators $\hat\pi_i$ for different families are made from the same finite-difference operator
\be
\hat\p_i := {1\over x_i-x_{i+1}}(1 - \sigma_i)
\ee
which satisfies
\be
\hat\p_i^2 = 0
\ee
and
\be
\hat \p_i \hat \p_{i+1} \hat\p_i = \hat \p_{i+1}\hat \p_i\hat\p_{i+1}
\ee

\bigskip

This input defines various families of non-symmetric polynomials \cite{Alex}:

\bigskip

\centerline{
\begin{tabular}{|c|c|c|}
\hline &&\\
family of polynomials & $\hat\pi_i^{family}$ & $\lambda$ \\ \hline
&&\\
Schubert polynomials &  $\hat\p_i$ & $\lambda_0$ \\
Key polynomials& $\hat \p_i x_i$ & any \\
Demazure atoms polynomials & $x_{i+1}\hat \p_i$ & any \\
Grothendieck polynomials& $\hat\p_i(1-x_{i+1})$   &$\lambda_0$ \\
$\ldots$ &&  \\
non-sym Macdonald polynomials& $T_i$ & any \\
$\ldots$ & &\\
\hline &&
\end{tabular}
}

\bigskip

Here $T_i$ is the Demazure-Lustig operator, see the next section.

It can be also instructive to compare $\hat\p_i$ with the Dunkl operators,
\be
\mathfrak{d}_i :=\frac{\p}{\p x_i} +\beta \sum_{j\neq i} \frac{1-\hat\sigma_{i,j}}{x_i-x_j}
\ee
which involve permutations at any distance, not only between the neighbours.

Acting on symmetric functions
of $x_1,\ldots,x_{n}$,
all operators $\hat \pi_i$ with $i=1,\ldots,n-1$ produce symmetric functions of the same variables.
However, $\hat \pi_n$ adds a variable $x_{n+1}$, not obligatory in a symmetric way.
For example, the time variables on the Miwa locus,
\be
p_k^{(n)} :=\sum_{i=1}^n x_i^k
\ee
are annihilated by the action of $\hat\p_1,\ldots,\hat\p_{n-1}$, but
\be
\hat\pi_n^{Schubert} p_k^{(n)}=
\hat\p_n p_k^{(n)}= \frac{x_n^k-x_{n+1}^k}{x_n-x_{n+1}}= S_{[k-1]}[x_n,x_{n+1}]
\ee
where the Schur polynomial $S_R$ at the r.h.s. with the Young diagram $R=[k-1]$ depends only on two $x$-variables
and can not be symmetric in all the $n+1$.

Likewise, $\hat\pi_i^{Key} =\hat\p_i x^i$ with $i<n$ leave all $p_k^{(n)}$ intact,
while
\be
\hat\pi_n^{key} p_k^{(n)}= p_k^{(n-1)}+\frac{x_n^{k+1}-x_{n+1}^{k+1}}{x_n-x_{n+1}}=
p_k^{(n-1)}+S_{k}[x_n,x_{n+1}]
\ee
which is no longer invariant under action of the permutation operators $\hat\sigma_{i,n}$ and $\hat\sigma_{i,n+1}$
with $i=1,\ldots,n-1$.
The situation is similar for other families.

Thus, the consecutive action of $\hat\pi$ gives rise to {\it non}-symmetric polynomials for two reasons:
original $x^\lambda$ can be asymmetric
(unless the diagram $\lambda$ is rectangular, $[r^n]$, and is invariant under permutations of $n$ variables),
but the asymmetry is continuously decreased by consecutive application of $\hat\pi_i$-operators with $i<n$,
or because of the action of operators $\hat\pi_n$ at the boundary,
which changes the number of $x$-variables and can break the symmetry, even if it was already achieved.
This second origin of the asymmetry can be eliminated by fixing $n$, and forbidding the application of $\hat\pi_n$.
 Then the symmetry is gradually increasing with the distance from the origin $x^\lambda$
 and, at some moment, the polynomials become fully symmetric.

In other words, different families of non-symmetric polynomials {\it stabilize} at the families of symmetric ones,
 where we primarily distinguish the Schur-Jack-Macdonald family. For details and an extensive list of sources, see \cite{Alex}.

\section{Macdonald non-symmetric polynomials \label{nsMac}}

In the remaining part of the paper, we first consider non-symmetric polynomials from the Macdonald family.
This system of polynomials is defined as a system of common eigenfunctions of the ordinary commuting Cherednik operators,
which are the finite-difference generalizations of Dunkl operators,
made with the help of ${\cal R}$-matrices.
These eigenfunctions are enumerated/labeled by \wcs, or by points of the integral weight lattice.
In the corresponding limits, the non-symmetric Macdonald polynomials give rise to the Demazure atoms and key polynomials.
At the same time, they preserve some similarities with the symmetric Macdonald polynomials,
enumerated and labeled by Young diagrams, or by dominant integer weights.
We review Macdonald theory in  sec.\ref{nsMac} in a way, not quite standard for traditional presentations in this field.

Our main interest is, however in another direction, which we discuss in the following sections.
It is about a further generalization to {\it twisted} functions,
which are the common eigenfunctions
of the cleverly designed $a$-th powers of Cherednik operators,
preserving their commutativity.
The main origin of our interest to twisting lies in application to representation
theory of the DIM algebra, where commuting families of operators along ``rays" $(b,a)$
form new families of integrable systems.
Pure twisting corresponds to a simpler family of ``integer rays" $(1,a)$.

These commuting families of operators in the $n$-body representation of the DIM algebra, when acting on the space of symmetric functions can be realized as power sums of a twisted version of the Cherednik operators. The twisted Cherednik operators are also commuting, and studying their eigenfunctions is our main target in secs.4-6. In the simplest untwisted case, these eigenfunctions are the non-symmetric Macdonald polynomials (sec.3), while, at generic $a$, all eigenfunctions are constructed as linear combinations of some basic ground state solutions in a universal way so that the coefficients of these combinations do not depend on the twist $a$ at all, and all the $a$-dependence is hidden in the ground state solutions. In its turn, these ground state solutions turns out to be the multivariable Baker-Akhiezer functions \cite{Cha}, which are eigenfunctions \cite{ChF,MMP1} of the commuting families of operators in the DIM algebra associated with the integer rays \cite{MMP}.

\subsection{Basic operators \label{bops}}

Cherednik operators $C_k$ and Demazure-Lustig operators $T_i$ are defined \cite{Ch,NSCh}\footnote{
In the limit $t\longrightarrow 0$ the operator
$$ T_i = \frac{(t^{-1}-1)x_{i+1}}{x_i-x_{i+1}} + \frac{x_i-t^{-1}x_{i+1}}{x_i-x_{i+1}}\sigma_{i,i+1}$$
becomes proportional to Demazure operator
$$ \hat \pi_i^{Demazure} = x_{i+1}\p_i = \frac{x_{i+1}}{x_i-x_{i+1}} - \frac{x_{i+1}}{x_i-x_{i+1}} \sigma_{i,i+1}$$
In the limit $t\longrightarrow \infty$, it turns into
$$\frac{x_{i+1}}{x_i-x_{i+1}} - \frac{x_i}{x_i-x_{i+1}} \sigma_{i,i+1}$$
which   differs by reordering from
$$ \hat \pi_i^{key} = \p_i x_i = \frac{x_i}{x_{i+1}-x_i} - \frac{x_{i+1}}{x_{i+1}-x_i} \underbrace{\sigma_{i+1,i}}_{=\sigma_{i,i+1}} $$
}
\be
R_{ij}:&=&1+{(1-t^{-1})x_j\over x_i-x_j}(1-\sigma_{i,j})\\
R_{ij}^{-1}&=&1+{(1-t)x_j\over x_i-x_j}(1-\sigma_{i,j})\nn\\
T_i:&=&R_{i,i+1}\sigma_{i,i+1}=\sigma_{i,i+1}+{(t^{-1}-1)x_{i+1}\over x_i-x_{i+1}}(1-\sigma_{i,i+1})=
1+{x_i-t^{-1}x_{i+1}\over x_i-x_{i+1}}(\sigma_{i,i+1}-1),\ \ \ \ \ i=1,\ldots,n-1\nn\\
T_i^{-1}&=&\sigma_{i,i+1}R_{i,i+1}^{-1}\nn\\
T_0:&=&1+{qx_n-t^{-1}x_{1}\over qx_n-x_{1}}(\sigma_{1,n}q^{\hat D_1-\hat D_n}-1)\nn\\
C_i &=& t^{1 - i} \left(\prod_{j = i + 1}^n R_{i,j}\right)
    q^{\hat D_i}
    \left(\prod_{j = 1}^{i - 1} R^{-1}_{j,i}\right)=T_iT_{i+1}\ldots T_{n-1}\sigma_{i,n}q^{\hat D_1}\sigma_{1,i}T_1^{-1}\ldots T_{i-1}^{-1}
    \nn
\label{ordefs}
\ee
where $\hat D_i:=x_i{\p\over\p x_i}$. The products in $C_i$ are obtained so that the smaller index stands to the left.

These quantities satisfy a set of relations:
\begin{itemize}
\item At $i=1,\ldots,n-1$ (Hecke algebra):
\be
(T_i-1)(T_i+t^{-1})&=&0\nn\\
\phantom{.}[T_i,T_j]&=&0,\ \ \ \ \ \ \ |i-j|\ge 2\\
T_iT_{i+1}T_i&=&T_{i+1}T_iT_{i+1}
\ee
\item At $i=1,\ldots,n-1$:
\be
tT_iC_{i+1}T_i=C_i
\ee
\item At $i,j=1,\ldots,n$:
\be
\phantom{.}[C_i,C_j]=0
\ee
\end{itemize}

\subsection{Orthogonality relations}

The non-symmetric Macdonald polynomials are
\be\label{nsM}
E_{\lambda}=x^\lambda+\sum_{\mu<\lambda}C_{\lambda\mu}x^\mu
\ee
where $\lambda$ is a \wc with $n$ parts (unordered, and some of the parts may be zero). If there are two \wcs, $\lambda$ and $\mu$, $\lambda>\mu$ if the ordered partition $\lambda^+>\mu^+$ (e.g., in accordance with the lexicographic order), and if the ordered partitions coincide, one compares the minimal length of permutations of the symmetric group ${\cal S}_n$ that allow one to make an ordered partition. The less is the length, the larger is \wc. In other words, the largest one is $\lambda^+$, and in the sum in (\ref{nsM}) all $\mu$ with $\mu^+=\lambda^+$ are present. The next smaller one is any one $\mu_i$ from the set of $\{\mu_i=\sigma_i(\mu^+)\}$, $i=1,\ldots,n-1$ given by a single elementary transposition. Such $\mu_i$ does not include into the sum only $\mu^+$ and all the elements of this set, etc. This is called {\it Bruhat order}.

One can use two ways to unambiguously restore the coefficients $C_{\lambda\mu}$ in (\ref{nsM}): there is an orthogonality condition with respect to the Cherednik scalar product:
\be\label{OR}
\Big<f,g\Big>=\prod_{i=1}^n\oint{dx_i\over x_i}f(x_i;q,t)g(x_i^{-1};q^{-1},t^{-1})\prod_{i>j}{(x_i/x_j;q)_\infty(qx_j/x_i;q)_\infty
\over (tx_i/x_j;q)_\infty(tqx_j/x_i;q)_\infty}
\ee
This scalar product does not look too effective for constructing the non-symmetric Macdonald polynomials because of necessity of making the replace $(q,t)\to (q^{-1},t^{-1})$ in the second polynomial.

\subsection{Non-symmetric Macdonald polynomials as eigenfunctions of the Cherednik operators}

The second way, which is quite effective, is to use that the Cherednik operators $C_i$ commute with each other, and their system of eigenfunctions is given by the non-symmetric Macdonald polynomials so that the coefficients $C_{\lambda\mu}$ in (\ref{nsM}) are fixed unambiguously. Thus, one solves the equations
\be\label{ChE}
C_i\cdot E_\lambda=\Lambda^{(i)}_\lambda E_\lambda,\ \ \ \ \ \ \ \ \ \ i=1,2,\ldots,n
\ee
where $\Lambda_\lambda^{(i)}$ are eigenvalues. If one considers solutions of a given homogeneity $p$ in $x_i$, these equations have the number of non-trivial solutions as many as the number of \wcs $\lambda$ of $p$ in $n$ parts, which are just the non-symmetric Macdonald polynomials $E_\lambda$. Note that, with the notation used here, the polynomials are obtained with opposite numeration of $x_i$ as compared with \cite{HHL}:
\be\label{Ens}
E_{[0,0,1]}&=&x_3\nn\\
E_{[0,1,0]}&=&x_2+{qt(1-t)\over 1-qt^2}x_3\nn\\
E_{[1,0,0]}&=&x_1+{q(1-t)\over 1-qt}(x_2+x_3)\nn\\
E_{[0,0,2]}&=&x_3^2+{1-t\over 1-qt}(x_1x_3+x_2x_3)\nn\\
E_{[0,2,0]}&=&x_2^2+{q^2t(1-t)\over (1-q^2t^2)}x_3^2+
{1-t\over 1-qt}x_1x_2+{q^2t(1-t)^2\over (1+qt)(1-qt)^2}x_1x_3+{q(1-t)(1+qt-qt^2-q^2t^2)\over (1+qt)(1-qt)^2}x_2x_3\nn\\
E_{[2,0,0]}&=&x_1^2+{q^2(1-t)\over 1-q^2t}(x_2^2+x_3^2)+{q(1+q)(1-t)\over 1-q^2t}(x_1x_2+x_1x_3)+{q^2(1+q)(1-t)^2\over
(1-qt)(1-q^2t)}x_2x_3\nn\\
E_{[0,1,1]}&=&x_2x_3\nn\\
E_{[1,0,1]}&=&x_1x_3+{qt(1-t)\over 1-qt^2}x_2x_3\nn\\
E_{[1,1,0]}&=&x_1x_2+{q(1-t)\over 1-qt}(x_1x_3+x_2x_3)
\ee
at $n=3$. One can immediately obtain the $n=2$ case at $x_3=0$: $E_{[\lambda_1,\lambda_2,0]}(x_1,x_2,x_3)\Big|_{x_3=0}=E_{[\lambda_1,\lambda_2]}$, and $E_{[\lambda_1,\lambda_2,\lambda_3]}(x_1,x_2,x_3)\Big|_{x_3=0}=0$ if $\lambda_3\ne 0$. This is the stability property of the non-symmetric Macdonald polynomials. In particular,
\be
E_{[0, 3]}&=& x_2^3 + {1-t\over 1-q^2t}x_1^2x_2 + {(1-t)(1+q)\over 1-q^2t}x_1x_2^2\nn\\
E_{[3, 0]}&=& x_1^3 + q^3{1-t\over 1-q^3t}x_2^3 + {q(1-t)(1+q+q^2)\over 1-q^3t}x_1^2x_2 + q^2{(1-t)(1-qt)(1+q+q^2)\over (1-q^2t)(1-q^3t)}x_1^2x_2\nn\\
E_{[1, 2]}&=& x_1x_2^2\nn\\
E_{[2, 1]}&=& x_1^2x_2 + q{1-t\over 1- qt}x_1x_2^2
\ee

The eigenvalues $\Lambda^{(i)}_\lambda$ are:
\be
\Lambda^{(i)}_\lambda=q^{\lambda_i}t^{-\zeta(\lambda)_i}
\ee
where $\zeta(\lambda)_i:=\#\{k<i|\lambda_k\ge\lambda_i\}+\#\{k>i|\lambda_k>\lambda_i\}$.

Moreover, solutions of equations (\ref{ChE}) has a natural triangle structure: requiring the unit coefficient in front of $x_1$ in the example above, one obtains the only solution $E_{[0,0,1]}$, requiring the unit coefficient in front of $x_2$, one obtains the an additional solution $E_{[0,1,0]}$, etc.

Note that the symmetric Macdonald polynomials can be similarly unambiguously (up to a normalization) obtained as solutions to the one eigenvalue equation
\be
H^{RC}_1\cdot M_{\lambda^+}=\bar\Lambda^{(k)}_{\lambda^+}M_{\lambda^+},
\ee
where $H^{RC}_1$ is the Macdonald-Ruijsenaars operator, which is the first of $n$ commuting Hamiltonians $H^{RC}_k$, $k=1..n$:
$[H^{RC}_k,H^{RC}_l]=0$.

Note that,when acting on the space of symmetric functions, these commuting Hamiltonians coincide with the power sums of the Cherednik operators, or, equivalently, they coincide with the symmetrized power sums
\be\label{HC}
H_k^{RC}=\sum_iC_i^k\Big|_{symm}=\hbox{Sym}\left(\sum_iC_i^k\right)
\ee
Hence, one can write
\be
\sum_iC_i^k\cdot M_{\lambda^+}=\bar\Lambda^{(k)}_{\lambda^+}M_{\lambda^+}
\ee
from (\ref{HC})
and
\be\label{HRCCE}
\sum_iC_i^k\cdot E_\lambda=\sum_i\Big(\Lambda^{(i)}_\lambda\Big)^k E_\lambda
\ee
from (\ref{ChE}). However, the later equations, (\ref{HRCCE}) do not fix non-symmetric solutions unambiguously. For instance, an arbitrary linear combination in $x_i$ solves (\ref{HRCCE}).

\subsection{Properties of non-symmetric Macdonald polynomials}

Note that, at $q=1$, when $\lambda$ is an ordered partition, $E_\lambda$ becomes a symmetric polynomial. Moreover, in general
$E_\lambda$ at $q=1$ factors into a symmetric and a non-symmetric parts, and the symmetric part is independent of $t$.
For instance,
\be
E_{[1,0,0]}\Big|_{q=1}=S_{[1]}\nn\\
E_{[1,1,0]}\Big|_{q=1}=S_{[11]}\nn\\
E_{[2,0,0]}\Big|_{q=1}=p_1^2=S_{[11]}+S_{[2]}\nn\\
E_{[3,0,0]}=p_1^3\nn\\
E_{[2,1,0]}=p_1S_{[1,1]}\nn\\
E_{[1,1,1]}=S_{[1,1,1]}\nn\\
E_{[0,2,0]}\Big|_{q=1}=p_1\Big(x_2+{t\over 1+t}x_3\Big)\nn\\
E_{[0,0,2]}\Big|_{q=1}=p_1x_3\nn\\
E_{[0,3,0]}=p_1^2\Big(x_2+{t\over 1+t}x_3\Big)\nn\\
E_{[0,0,3]}=p_1^2x_3\nn\\
E_{[2,0,1]}=p_1\Big(x_1+{t\over 1+t}x_2\Big)x_3\nn\\
\ee

Note also that there a symmetry
\be
E_{[0,1,1]}(x_1,x_2,x_3)&=&x_3E_{[0,0,1]}(q^{-1}x_3,x_1,x_2)\nn\\
E_{[1,0,1]}(x_1,x_2,x_3)&=&x_3E_{[0,1,0]}(q^{-1}x_3,x_1,x_2)\nn\\
E_{[0,0,2]}(x_1,x_2,x_3)&=&qx_3E_{[1,0,0]}(q^{-1}x_3,x_1,x_2)
\ee
These are particular cases of the general identity
\be\label{sym2}
E_{[\lambda_2,\ldots,\lambda_n,\lambda_1+1]}(x_1,x_2,\ldots,x_n)=
q^{\lambda_1}x_nE_{[\lambda_1\lambda_2,\ldots,\lambda_n]}(q^{-1}x_{n},x_1,x_2,\ldots,x_{n-1})
\ee

\subsection{Creation operators}

Note that operators $T_i$ allows one to construct the non-symmetric Macdonald polynomials recursively. Indeed, the action of this operators just permutes the $i$-th and $(i+1)$-th parts of the \wc so that
\be
T_iE_\lambda&=&E_\lambda,\ \ \ \ \ \ \hbox{if}\ \ \ \ \ \lambda_i=\lambda_{i+1}\nn\\
T_iE_\lambda&=&\alpha_{i,\lambda}E_\lambda+E_{\sigma_i\lambda},\ \ \ \ \ \ \hbox{if}\ \ \ \ \ \lambda_i<\lambda_{i+1}\nn\\
T_iE_\lambda&=&\alpha_{i,\lambda}E_\lambda+\beta_{i,\lambda} E_{\sigma_i\lambda},\ \ \ \ \ \ \hbox{if}\ \ \ \ \ \lambda_i>\lambda_{i+1}
\ee
where $\sigma_i\lambda$ permutes the $i$-th and $(i+1)$-th parts of $\lambda$, and $\alpha_i$, $\beta_i$ are some constants of $q$ and $t$:
\be
\alpha_{i,\lambda}:&=&-{(1-t)\over t(1-A_i^{-1})}\nn\\
\beta_i:&=&{( 1-A_it)( 1-A_i/t)\over t( 1-A_i)^2}
\ee
where
\be
A_i:=q^{\lambda_{i}-\lambda_{i+1}}t^{\zeta(\lambda)_{i+1}-\zeta(\lambda)_{i}}
\ee
For instance:
\be
T_1E_{[0,0,1]}&=&E_{[0,0,1]}\nn\\
T_1E_{[0,1,0]}&=&-{1-t\over t(1-qt)}E_{[0,1,0]}+E_{[1,0,0]}\nn\\
T_1E_{[1,0,0]}&=&-{1-t\over t(1-q^{-1}t^{-1})}E_{[1,0,0]}+{(1-q)(1-qt^2)\over t(1-qt)^2}E_{[0,1,0]}\nn\\
T_2E_{[0,0,1]}&=&-{1-t\over t(1-qt^2)}E_{[0,0,1]}+E_{[0,1,0]}\nn\\
T_2E_{[0,1,0]}&=&-{1-t\over t(1-q^{-1}t^{-2})}E_{[0,1,0]}+{(1-qt)(1-qt^3)\over t(1-qt^2)^2}E_{[0,0,1]}\nn\\
T_2E_{[1,0,0]}&=&E_{[1,0,0]}
\ee

Another important property is the stability: $E_\lambda(x_1,\ldots,x_{n-1},0)=0$ if $\lambda_n\ne 0$, and $E_\lambda(x_1,\ldots,x_{n-1},0)=E_{\lambda'}(x_1,\ldots,x_{n-1})$ otherwise, where $\lambda'$ denotes the $n$-th (zero) part removed.

So far, we had operators that permuted parts of the \wc $\lambda$. Now we construct the operator that increases \wcs:
\be
\hat B:=x_nT_{n-1}^{-1}T_{n-2}^{-1}\ldots T_{2}^{-1}T_{1}^{-1}
\ee
It acts on the non-symmetric Macdonald polynomials in the following way:
\be
\hat B\cdot E_{[\lambda_1,\ldots,\lambda_n]}=t^{n-1-\#\{\lambda_i\le\lambda_1\}}E_{[\lambda_2,\ldots,\lambda_{n},\lambda_1+1]}
\ee
In fact, this operator uses the symmetry (\ref{sym2}).

\bigskip

Though carrying the same name,
these creation operators are substantially distinct from Kirillov-Noumi ones \cite{KN}, reviewed recently in \cite{MMkn}.

\subsection{Cauchy identity}

The Cauchy identity for the non-symmetric Macdonald polynomials looks as
\be\label{CI}
\sum_{\lambda\in\mathbb{Z}^n_+}a_\lambda(q,t)E_\lambda(x;q,t)E_\lambda(y;q^{-1},t^{-1})=\exp\left(\sum_k{1-t^k\over 1-q^k}{p_k\bar p_k
\over k}\right)\prod_{i=1}^n{1\over 1-tx_iy_i}\prod_{i>j}{1-x_iy_j\over 1-tx_iy_j}
\ee
where $p_k:=\sum_i^nx_i^k$, $\bar p_k:=\sum_i^ny_i^k$, and
\be
a_\lambda(q,t):=\prod_{s=(i,j)\in\lambda}{1-q^{a(s)+1}t^{l(s)+1}\over 1-q^{a(s)+1}t^{l(s)}}
\ee
with $a(s)=\lambda_i-j$ being the standard arm length, while the leg length $l(s)$ is defined as the number of $k>i$: $j\le \lambda_k\le\lambda_i$ plus the number of $k<i$: $j\le \lambda_k+1\le\lambda_i$. Such defined leg length coincides with the standard one when $\lambda=\lambda^+$.

Note that the sum in formula (\ref{CI}) involves both the non-symmetric Macdonald polynomials at $(q,t)$ and $(q^{-1},t^{-1})$. This is not surprising because the orthogonality relation (\ref{OR}) also involves both of these points, and the Cauchy identity is related to the orthogonality relation \cite{CO}.

\subsection{Limit to non-symmetric Jack polynomials}

One can take the limit from the construction of the previous section and obtain, as counterparts of the Cherednik operators, the operators of the form:
\be\label{Dl}
{\cal D}_i:=x_i{\p\over\p x_i}+\beta\sum_{i\ne j}{x_i\over x_i-x_j}(1-\sigma_{ij})+\beta\sum_{j>i}\sigma_{ij}
=x_i\mathfrak{d}_i+\beta\sum_{j>i}\sigma_{ij}
\ee
where $\mathfrak{d}_i$ is the Dunkl operator. The operators ${\cal D}_i$ are commuting, and the system of their eigenvalues is nothing but the non-symmetric Jack polynomials, the first of them being
\be
J_{[0,0,1]}&=&x_3\nn\\
J_{[0,1,0]}&=&x_2+{\beta\over 2\beta+1}x_3\nn\\
J_{[1,0,0]}&=&x_1+{\beta\over \beta+1}(x_2+x_3)
\ee
This can be naturally obtained from the non-symmetric Macdonald polynomials with the parametrization $t=q^\beta$ in the limit of $q\to 1$.

One also can naturally associate the $\beta=1$ case with the non-symmetric Schur polynomials. In particular, these non-symmetric Schur functions satisfy the Cauchy identity (\ref{CI}) that involves only these Schur functions.

\subsection{Limit to Demazure atoms and key polynomials}

In fact, there are three natural choices of the Schur limits: $q=t\to 0,1,\infty$. In the case of $q=t\to\infty$, one obtains the key polynomials:
\be
E_\lambda(x_1,\ldots,x_n;\infty,\infty)=K_{w_0\lambda}(x_n,\ldots,x_1)
\ee
where $w_0$ is the longest permutation in permutation group ${\cal S}_n$,
and, in the case of $q=t\to 0$, the Demazure atoms:
\be
E_\lambda(x_1,\ldots,x_n;0,0)=A_\lambda(x_1,\ldots,x_n)
\ee
These two kinds of non-symmetric polynomials are both involved into the corresponding Cauchy identity \cite[Theorem 6]{Las03},\cite{Alex}.

Now let us put in the Cauchy identity (\ref{CI}) $q=t=0$. Then, one immediately obtains
\be
\sum_\lambda E_\lambda(x;0,0)E_\lambda(y;\infty,\infty)==\prod_{i\le j}{1\over 1-x_iy_j}
\ee
In order to compare this formula with \cite[Theorem 6]{Las03},\cite{Alex}, notice the inverse order of $x_i\to x_{n-i+1}$ in the key polynomials.

Let us see how this identity works. The first Demazure atoms are
\be
E_{[1,0,0]}(x;0,0)=x_1\nn\\
E_{[0,1,0]}(x;0,0)=x_2\nn\\
E_{[0,0,1]}(x;0,0)=x_3
\ee
and the key polynomials are
\be
E_{[1,0,0]}(y;\infty,\infty)&=&y_1+y_2+y_3\nn\\
E_{[0,1,0]}(y;\infty,\infty)&=&y_2+y_3\nn\\
E_{[0,0,1]}(y;\infty,\infty)&=&y_3
\ee
Then,
\be
\sum_\lambda E_\lambda(x;0,0)E_\lambda(y;\infty,\infty)=1+x_1y_1+x_1y_2+x_1y_3+x_2y_2+x_2+y_3+x_3y_3+\ldots
\ee
which is equal to the linear terms of expansion of
\be
\prod_{i\le j\le 3}{1\over 1-x_iy_j}={1\over (1-x_1y_1)(1-x_1y_2)(1-x_1y_3)(1-x_2y_2)(1-x_2y_3)(1-x_3y_3)}=\nn\\
=1+x_1y_1+x_1y_2+x_1y_3+x_2y_2+x_2+y_3+x_3y_3+\ldots
\ee

\subsection{Symmetric Macdonald polynomials}

Symmetric Macdonald polynomials associated with the dominant integral weights can be obtained from the non-symmetric Macdonald polynomials by summing up over the Weyl group $W={\cal S}_n$, i.e. over all permutations of the partition $\lambda^+$:
\be\label{nss}
M_{\lambda^+}=\sum_{{\lambda=w\cdot\lambda^+}\atop{w\in W}} E_\lambda\cdot\left(\prod_{(i,j):\ \lambda_j>\lambda_i}{1-q^{\lambda_j-\lambda_i}t^{\zeta(\lambda)_i-\zeta(\lambda)_j-1}\over
1-q^{\lambda_j-\lambda_i}t^{\zeta(\lambda)_i-\zeta(\lambda)_j}}\right)
\ee
The product in the summand runs over pairs of $(i,j)$ such that $\lambda_i<\lambda_j$. This gives the symmetric Macdonald polynomials in the standard normalization of the $P$ polynomials \cite{Mac}.

\section{Integrable systems associated with DIM algebra\label{intDIM}}

\subsection{Commutative subalgebras of DIM algebra}

We are going to realize the construction of sec.2.1 in the case when the operators $\hat h_k$ are the Hamiltonians of integrable systems associated with integer rays of the DIM algebra \cite{MMP}. We use the elliptic Hall algebra formulation of the DIM algebra. The elliptic Hall algebra is an associative algebra  multiplicatively generated by two central elements and elements $\mathfrak{e}_{\vec{\gamma}}$, with $\vec{\gamma} \in \mathbb{Z}^2\setminus \{(0,0)\}$, satisfying a set of commutation relations \cite{Feigin,BS,Zenk}. An important property of this algebra is that any vector $\vec\gamma$ gives rise to a commutative subalgebra:
\be\label{sb}
\left[ \mathfrak{e}_{\vec\gamma}, \mathfrak{e}_{k\vec \gamma}\right] = 0 \ \ \ \ \ \forall \vec\gamma \ {\rm and} \ k\in Z_+
\ee
The subalgebras associated with rays $\mathfrak{e}_{(\pm 1,a)}$ are called integer rays \cite{MMP}. In fact, all these subalgebras are related by the Miki automorphisms \cite{Miki1}, which represent action of the $SL(2,\mathbb{Z})$ group.

Various representations of the DIM algebra have been studied, we will concentrate on the $n$-body (or $n$-particle) representation of the algebra \cite{MMP}, which is just a tensor power of the vector representation \cite{Zenk2}. Commutative subalgebras in this representation give rise to integrable Hamiltonians of many-body systems, which generalize the trigonometric Ruijsenaars-Schneider systems.

We will discuss only integer rays $\mathfrak{e}_{(-1,a)}$, since the reflection symmetries: $\mathfrak{e}_{(k,m)}(x;q,t)\sim \mathfrak{e}_{(-k,m)}(x^{-1};q^{-1},t^{-1})$ and $\mathfrak{e}_{(k,m)}(q,t)=-\mathfrak{e}_{(k,-m)}(q^{-1},t^{-1})$ relate the rays in different quadrant of the $2d$ integer plane.

\subsection{Hamiltonians in $n$-body representation through higher Cherednik operators}

In the $n$-body representation, the commutative subalgebra associated with ray $\mathfrak{e}_{(0,1)}$ is just a set of trigonometric Ruijsenaars-Schneider Hamiltonians. They can be rewritten in the form (\ref{HC}) when acting on the space of symmetric functions:
\be
H_k^{RC}=\sum_iC_i^k\Big|_{symm}=\hbox{Sym}\left(\sum_iC_i^k\right)
\ee
Here we have another example of the construction of sec.2.1 with operators $\hat c_i$ corresponding to the Cherednik operators $C_i$, and the operators $\hat h_k$ corresponding to the Ruijsenaars-Schneider Hamiltonians.

Another commutative subalgebra, the one associated with the ray $\mathfrak{e}_{(-1,1)}$, i.e. consisting of elements $\mathfrak{e}_{[-k,k]}$ is given by the operators
\be\label{calCi}
{\cal C}_i=C_i\Big|_{q^{\hat D_i}\to{1\over x_i}q^{\hat D_i}}=
t^{1 - i} \left(\prod_{j = i + 1}^n R_{i,j}\right)
    {1\over x_i}q^{\hat D_i}
    \left(\prod_{j = 1}^{i - 1} R^{-1}_{j,i}\right)
\ee
instead of $\hat c_i$ and Hamiltonians
\be
H_k^{(1)}:=\sum_i{\cal C}_i^k\Big|_{symm}=\hbox{Sym}\left(\sum_i{\cal C}_i^k\right)
\ee
instead of $\hat h_k$.

Similarly, higher commutative subalgebras associated with the ray $\mathfrak{e}_{(-1,a)}$, i.e. consisting of elements $\mathfrak{e}_{[-k,ka]}$ are given by the higher Cherednik operators
\be\label{aC}
\mathfrak{C}_i^{(a)}:={1\over x_i}\Big(x_i{\cal C}_i\Big)^a
\ee
instead of $\hat c_i$, and Hamiltonians
\be
H_k^{(a)}:=\sum_i\left(\mathfrak{C}_i^{(a)}\right)^k\Big|_{symm}=\hbox{Sym}\left(\sum_i\left(\mathfrak{C}_i^{(a)}\right)^k\right)
\ee
instead of $\hat h_k$.

\subsection{Eigenfunctions: twisted Baker-Akhiezer functions}

The eigenfunctions that are counterparts of $e_\lambda$ and $s_{\lambda^+}$ for the ray $\mathfrak{e}_{(0,1)}$ are nothing but the non-symmetric Macdonald polynomials $E_\lambda$ and symmetric Macdonald polynomials $M_{\lambda^+}$. More interesting are the eigenfunctions for the rays $\mathfrak{e}_{(-1,a)}$. Here we discuss eigenfunctions of the Hamiltonians $H_k^{(a)}$, i.e. counterparts of $s_{\lambda^+}$, and, in the next sections, we construct eigenfunctions of ${\cal C}_i^{(a)}$, i.e. counterparts of $e_{\lambda}$.

The simplest case is the ray $\mathfrak{e}_{(-1,1)}$ when the corresponding eigenvalue equation reads
\be
\hat H^{(1)}_k\left[q^{{1\over 2}\sum_iz_i^2}\cdot M_{\lambda^+}\right]=t^{k\over 2}\left(\sum_iq^{k\lambda_i}\right) \left[q^{{1\over 2}\sum_iz_i^2}\cdot M_{\lambda^+}\right]
\ee
and we denoted $x_i=q^{z_i}$.

In order to deal with the case of other $\mathfrak{e}_{(-1,a)}$ rays, we restrict ourselves with values of $t=q^{-m}$ with integer $m$. Then, solution is \cite{MMP1} the so called twisted Baker-Akhiezer function \cite{Cha,ChE}, which is non-symmetric (quasi)polynomial but of a distinct type as compared with non-symmetric polynomials considered above. The twisted BA function, which is a function of $2n$ complex parameters $x_i=q^{z_i}$ and $y_i=q^{\lambda_i}$, $i=1,\ldots,n$, and is defined as a sum
\be\label{BAtN}
\Psi_m^{(a)}(\vec z,\vec\lambda)=q^{{\vec\lambda\cdot\vec z\over a}+m\vec\rho\cdot \vec z}\ \sum_{k_{ij}=0}^{ma}q^{-\sum_{i>j}{k_{ij}\over a}(z_i-z_j)}\psi^{(a)}_{m,\vec\lambda,k}
\ee
with the property
\be
\Psi_m^{(a)}(z_k+j,\vec\lambda)=\varepsilon^j\Psi_m^{(a)}(z_l+j,\vec\lambda)\ \ \ \ \  \forall k,l\ \ \hbox{and}\ \ 1\le j\le m\ \ \ \ \ \hbox{at}\ \ \varepsilon q^{z_k\over a}= q^{z_l\over a}
\ee
for any $\varepsilon$ such that $\varepsilon^a=1$. Here $\vec\rho$ is the Weyl vector, i.e. $\vec\rho\cdot\vec z={1\over 2}\sum_{i=1}^N(N-2i+1)z_i$. This $a$-twisted Baker-Akhiezer function is unique up to a normalization, and, upon a proper normalization, is symmetric with respect to permutation of $\vec x$ and $\vec\lambda$.

In the case of $a=1$, a proper sum of the Baker-Akhiezer function over the permutations of $x_i$ gives rise to the symmetric Macdonald polynomial. Moreover, in this case, the Baker-Akhiezer function is also the eigenfunction of the Ruijsenaars-Schneider Hamiltonians $H_k^{RC}$ and
\be
\hat H_k^{(1)}\left[q^{{1\over 2}\sum_iz_i^2}\cdot\Psi_m(\vec z,\vec \lambda)\right]=q^{-{km\over 2}}\left(\sum_iq^{k\lambda_i}\right) \left[q^{{1\over 2}\sum_iz_i^2}\cdot\Psi_m(\vec z,\vec \lambda)\right]
\ee
Similarly, for arbitrary $a$,
\be
\boxed{
\hat H^{(a)}_k\left[q^{{1\over 2a}\sum_iz_i^2}\cdot\Psi_m^{(a)}(z,\lambda)\right]=
q^{-{amk\over 2}}\left(\sum_iq^{k \lambda_i}\right)\left[ q^{{1\over 2a}\sum_iz_i^2}\cdot\Psi_m^{(a)}(z,\lambda)\right]
}
\ee

\section{Twisted Cherednik integrable systems\label{intCh}}

\subsection{Eigenfunctions of higher Cherednik Hamiltonians}

Now we discuss the counterparts of $e_\alpha$ in the twisted case. As we explained, the Cherednik operators $C_i$ are associated with the Ruijsenaars integrable Hamiltonians, i.e. with the commutative subalgebra of the elliptic Hall (DIM) algebra consisting of the elements $e_{[0,k]}$. Another commutative subalgebra consisting of elements $e_{[-k,k]}$ is associated with the operators ${\cal C}_i$ in (\ref{calCi}). The eigenfunctions of these operators form another set of non-symmetric functions, while their power sums give rise to symmetric functions proportional to the Macdonald Hamiltonians. Solutions to the equations
\be
{\cal C}_i\cdot \Phi^{(1)}_\lambda=\Lambda^{(1,i)}_\lambda\cdot \Phi^{(1)}_\lambda,\ \ \ \ \ i=1,\ldots, n
\ee
are again labeled by \wcs $\lambda$. Moreover, the solutions turns out to be proportional to the non-symmetric Macdonald polynomials:
\be
\Phi^{(1)}_\lambda=q^{{1\over 2}\sum_{i=1}^nz_i^2}\cdot E_\lambda
\ee
where we denoted $x_i=q^{z_i}$.

Higher commutative subalgebras are associated with the higher Cherednik Hamiltonians $\mathfrak{C}_i^{(a)}$ in (\ref{aC}),
and the equations
\be\label{Cev}
\mathfrak{C}_i^{(a)}\cdot \Phi^{(a)}_\lambda=\Lambda^{(a,i)}_\lambda\cdot \Phi^{(a)}_\lambda
\ee
have solutions of the form
\be\label{Cm}
\Phi^{(a)}_\lambda=q^{{1\over 2a}\sum_{i=1}^nz_i^2}\cdot  \psi^{(a)}_\lambda
\ee
where $\psi^{(a)}_\lambda$ at $a>1$ are some new functions, which are non-symmetric functions of $x^{1\over a}_i$ so that they can be naturally called twisted non-symmetric Macdonald functions. They become polynomials at $t=q^{-m}$, $m\in \mathbb{N}$. In practice, the multiplication by this factor means that we substitute $x_i^{-1}$ in front of dilatation $\frac{1}{x_i}q^{\hat D_i}$ within
Cherednik operators by $x_i^{\frac{1}{a}-1}$.

Note that any solution $\psi^{(a)}_\lambda$ can be multiplied by $\prod_{i=1}^nx_i^\alpha$ with arbitrary $\alpha$ still remaining a solution, since
\be\label{fac}
{\cal C}_i\Big(\prod_{i=1}^nx_i^\alpha\cdot F(x)\Big)=q^\alpha \prod_{i=1}^nx_i^\alpha\cdot{\cal C}_iF(x)
\ee

\subsection{Basis eigenfunctions at $n=2$, $a=2$\label{5.2}}

First of all, we find two eigenfunctions that allow us to construct all other solutions.
Put $\boxed{t=q^{-m}}$, and consider $n=2$, $a=2$. Then, at $m=0$,
\be
\psi^{(2)}=x_1^{\lambda_1}x_2^{\lambda_2},\ \ \ \ \ \ \Lambda^{(2,1)}=q^{2\lambda_1+1},\ \ \ \ \ \ \Lambda^{(2,2)}=q^{2\lambda_2+1}
\ee
for any $\lambda_1$ and $\lambda_2$ (not obligatory integer).

At natural $m$, there are polynomial solutions of the form of the monomial prefactor $(x_1x_2)^\alpha$ multiplied, in accordance with (\ref{fac}), with a polynomial of $x_1^{1/2}$, $x_2^{1/2}$, and this prefactor only shifts the eigenvalues.
The solution always contains $m+1$ terms.
One can immediately find two solutions:
\be
\!\!
\boxed{
\begin{array}{rcl}
\psi_1^{(2)}(m;\alpha)&=&x_1^\alpha x_2^{\alpha}\cdot\sum_k\binom{m}{k}_q\cdot(q^{-k}x_1)^{m-k\over 2}x_2^{k\over 2}=
x_1^\alpha x_2^{\alpha}\cdot \prod_{j=0}^{m-1}\Big(\sqrt{x_1}+q^{j-\frac{m-1}{2}}\sqrt{x_2} \Big):=
x_1^\alpha x_2^{\alpha}\Omega(m)\\ \\
\Lambda^{(2,1)}&=&q^{2\alpha+m+1},\ \ \ \ \ \ \  \Lambda^{(2,2)}=q^{2(\alpha+m)+1}\\
\\
\psi_2^{(2)}(m;\alpha)&=&x_1^\alpha x_2^{\alpha+{1\over 2}}\cdot\sum_k\binom{m}{k}_q\cdot(q^{-(k+1)}x_1)^{m-k\over 2}x_2^{k+1\over 2}=\\
\\
&=&x_1^\alpha x_2^{\alpha+\frac{1}{2}}\cdot \prod_{j=0}^{m-1}\Big(q^{-\frac{1}{2}}\sqrt{x_1}+q^{j-\frac{m-1}{2}}\sqrt{x_2} \Big)
:=x_1^\alpha x_2^{\alpha+{1\over 2}}\bar\Omega(m)\\ \\
\Lambda^{(2,1)}&=&q^{2(\alpha+m)+1},\ \  \Lambda^{(2,2)}=q^{2\alpha+m+2}
\end{array}
}
\
\ee
where $\binom{m}{k}_q$ denotes the $q$-binomial coefficients.

Note that
\be\label{Omega12}
\bar\Omega(m;x_1,x_2)\sim\Omega(m;x_1,qx_2)
\ee

Note also that one can continue these solutions to an arbitrary $t$ in the form
\be\label{Omegat}
\psi_1^{(2)}(q,t;\alpha)&=&t^{-{1\over 2}z_1}x_1^\alpha x_2^{\alpha}{\Big(-\sqrt{qtx_2\over x_1};q\Big)_\infty
\over \Big(-\sqrt{qx_2\over tx_1};q\Big)_\infty}=x_1^\alpha x_2^{\alpha}\Omega(q,t;x_1,x_2),\ \ \ \ \ \ \ \Lambda^{(2,1)}=q^{2\alpha+1}t^{-1},\ \ \ \ \ \ \  \Lambda^{(2,2)}=q^{2\alpha+1}t^{-2}
\nn\\
\psi_2^{(2)}(q,t;\alpha)&=&t^{1\over 2}t^{-{1\over 2}z_1}x_1^\alpha x_2^{\alpha+{1\over 2}}{\Big(-\sqrt{q^2tx_2\over x_1};q\Big)_\infty
\over \Big(-\sqrt{q^2x_2\over tx_1};q\Big)_\infty}=x_1^\alpha x_2^{\alpha+{1\over 2}}\bar\Omega(q,t;x_1,x_2),\ \ \ \ \ \ \ \Lambda^{(2,1)}=q^{2\alpha+1}t^{-2},\ \ \ \ \ \ \  \Lambda^{(2,2)}=q^{2\alpha+2}t^{-1}\nn\\
\ee

Note that the first of these eigenfunctions is a kind of ground state, is symmetric (as should be the ground state) and, hence, is simultaneously an eigenfunction of the both twisted Macdonald and $\mathfrak{C}_i^{(p)}$ Hamiltonians. In order to see that this is, indeed, the case, we note that $\psi_1^{(2)}$ is proportional to the multivariable Baker-Akhiezer function \cite{Cha,ChE}, which is an eigenfunction of the twisted Macdonald Hamiltonians \cite{ChF,MMP1,MMPf}.

Indeed, one can check that the 2-twisted Baker-Akhiezer function at $n=2$ \cite{MMP1,MMPdet}, $\Psi_m(\lambda_1,\lambda_2;x_1,x_2)$,  satisfies the identity
\be\label{EBA}
\boxed{
\psi_1^{(2)}\Big(m;\alpha\Big)\sim \Psi_m^{(2)}\Big(2\alpha,2\alpha+m;x_1,x_2\Big)
}
\ee
and there are also additional ``superfluous" relations (at shifted $m$)
\be
\boxed{
\psi_1^{(2)}\Big(m+1;\alpha\Big)\sim \Psi_m^{(2)}\Big(2\alpha+1,2\alpha+m;x_1,x_2\Big)
}
\ee
An important point here is that the Baker-Akhiezer function is a quasipolynomial, and admits quasipolynomial extension to arbitrary $t$ \cite[Eq.(48)]{MMP1}, which is different from (\ref{Omegat}).

\subsection{Constructing polynomial eigenfunctions}

The main problem with constructing polynomial eigenfunctions is that the operators $\mathfrak{C}_i^{(2)}$ maps polynomials onto polynomials of higher degree. Rotating $\mathfrak{C}_i^{(2)}$ with $q^{{1\over 4}\sum_{i=1}^nz_i^2}$, one provides operators that do not change the grading, but makes rational functions from polynomials. What one can do is to additionally rotate with $\Omega$, $\bar \Omega$: $\mathfrak{C}_i^{(2)}\longrightarrow U_{1,2}^{-1}\mathfrak{C}_i^{(2)}U_{1,2}$, where $U_{1}:=q^{{1\over 4}\sum_{i=1}^nz_i^2}\Omega$, $U_{2}:=q^{{1\over 4}\sum_{i=1}^nz_i^2}\bar\Omega$. That is,
\be\label{O}
\hat {\cal O}_1:=U_1^{-1}\mathfrak{C}_1^{(2)}U_1={q^{3\over 2}\over t^{1\over 2}}\left[\Big(t^{-1}-1)x_2\sigma_1
+\Big({x_2\over t}-x_1\Big)\right]{\Big(\sqrt{qx_1\over t}+\sqrt{x_2}\Big)\Big(\sqrt{x_2\over qt}-\sqrt{x_1}\Big)\over(qx_1-x_2)(x_1-x_2)}q^{2\hat D_1}-\nn\\
-{q^{3\over 2}\over t^{1\over 2}}\Big(t^{-1}-1\Big)\sqrt{x_1x_2}
\left[\Big(t^{-1}-1\Big)x_2+\Big({x_2\over t}-x_1\Big)\sigma_1\right]{1\over(qx_2-x_1)(x_2-x_1)}q^{\hat D_1+\hat D_2}=\nn\\
={q^{3\over 2}\over t^{1\over 2}}\left[\Big(t^{-1}-1)x_2\sigma_1
+\Big({x_2\over t}-x_1\Big)\right]{\left[
\Big(\sqrt{qx_1\over t}+\sqrt{x_2}\Big)\Big(\sqrt{x_2\over qt}-\sqrt{x_1}\Big)q^{2\hat D_1}-
\Big(t^{-1}-1\Big)\sqrt{x_1x_2} q^{\hat D_1+\hat D_2}\sigma_1
\right]\over (qx_1-x_2)(x_1-x_2)}
\nn\\
\hat {\cal O}'_1:=U_2^{-1}\mathfrak{C}_1^{(2)}U_2={q^{3\over 2}\over t^{1\over 2}}{(t^{-1}-1)x_2\over(x_1-x_2)(x_1-qx_2)}
\left[q^{1\over 2}\Big({x_1\over qt}-x_2\Big)q^{2\hat D_2}\sigma_1-(t^{-1}-1)\sqrt{x_1x_2}q^{\hat D_1+\hat D_2}\right]+\nn\\
+{q^{3\over 2}\over t^{1\over 2}}{\Big(\sqrt{x_1\over t}+\sqrt{x_2}\Big)\Big(\sqrt{x_2\over t}-\sqrt{x_1}\Big)\over
(x_1-x_2)(qx_1-x_2)}\left[q^{1\over 2}\Big({x_2\over qt}-x_1\Big)q^{2\hat D_1}-(t^{-1}-1)\sqrt{x_1x_2}q^{\hat D_1+\hat D_2}\sigma_1\right]=
\nn\\
={q^{3\over 2}\over t^{1\over 2}}\left[(t^{-1}-1)x_2\sigma_1+
\left(\sqrt{x_1\over t}+\sqrt{x_2}\right)\left(\sqrt{x_2\over t}
-\sqrt{x_1}\right)\right]{\left[q^{1\over 2}\Big({x_2\over qt}-x_1\Big)q^{2\hat D_1}-
(t^{-1}-1)\sqrt{x_1x_2}q^{\hat D_1+\hat D_2}\sigma_1\right]\over(qx_1-x_2)(x_1-x_2)}
\nn\\
\hat {\cal O}_2:=U_1^{-1}\mathfrak{C}_2^{(2)}U_1={q^{7\over 2}\over t^{-{1\over 2}}}{\left(\sqrt{x_1\over qt}-\sqrt{x_2}\right)
\left(\sqrt{x_2\over qt}+\sqrt{x_1}\right)
\over(x_1-qx_2)(x_1-q^2x_2)}
\left(\Big({x_1\over q^2t}-x_2\Big)+x_2(1-t^{-1})\sigma_1\right)q^{2\hat D_2}+\nn\\
+{q^{3\over 2}\over t^{1\over 2}}{(1-t^{-1})x_1^{1\over 2}x_2^{1\over 2}
\over(x_1-x_2)(x_1-qx_2)}\Big((1-t^{-1})x_1-(x_1-t^{-1}x_2)\sigma_1\Big)q^{\hat D_1+\hat D_2}\nn\\
\hat {\cal O}'_2:=U_2^{-1}\mathfrak{C}_2^{(2)}U_2={q^3\over t^{1\over 2}}{{x_1\over qt}-x_2\over (x_1-qx_2)(x_1-q^2x_2)}
\left[\left(\sqrt{q^2x_2\over t}+\sqrt{x_1}\right)\left(\sqrt{x_1\over q^2t}-\sqrt{x_2}\right)q^{2\hat D_2}
-qx_2q^{2\hat D_2}\sigma_1\right]+\nn\\
+{q^{3\over 2}\over t^{1\over 2}}{(t^{-1}-1)\sqrt{x_1x_2}\over(x_1-x_2)(x_1-qx_2)}
\left[\left(\sqrt{x_1\over t}+\sqrt{x_2}\right)\left(\sqrt{x_2\over t}-\sqrt{x_1}\right)q^{\hat D_1+\hat D_2}\sigma_1-
(t^{-1}-1)x_1q^{\hat D_1+\hat D_2}\right]
\ee
where $\sigma_1$ within the square brackets act only to the right (not act to the denominator).

These operators already have polynomial eigenfunctions. The reason is that the operators $\hat {\cal O}_1$, $\hat {\cal O}_2$ maps integer grading polynomials of $x_1^{1\over 2}$, $x_2^{1\over 2}$ onto similar polynomials preserving grading. Similarly, the operators $\hat {\cal O}'_1$, $\hat {\cal O}'_2$ maps polynomials of half-integer (non-integer) grading onto similar polynomials preserving grading.

Now one can solve the equations
\be
\hat {\cal O}_1\cdot E_\lambda^{(2)}(x_1^{1\over 2},x_2^{1\over 2})=\Lambda^{(2,1)}\cdot E_\lambda^{(2)}(x_1^{1\over 2},x_2^{1\over 2})\nn\\
\hat {\cal O}_2\cdot E_\lambda^{(2)}(x_1^{1\over 2},x_2^{1\over 2})=\Lambda^{(2,2)}\cdot E_\lambda^{(2)}(x_1^{1\over 2},x_2^{1\over 2})
\ee
with the anzatz
\be
E_\lambda^{(2)}(x_1,x_2)=x^\lambda+\sum_{\mu<\lambda}C_{\lambda\mu}^{(2)}x^\mu
\ee
and $|\lambda|$ even, and realize that solutions are again numbered by \wcs!

Similarly, one solves the equations
\be
\hat {\cal O}'_1\cdot E_\lambda^{(2)}(x_1^{1\over 2},x_2^{1\over 2})=\Lambda^{(2,1)}\cdot E_\lambda^{(2)}(x_1^{1\over 2},x_2^{1\over 2})\nn\\
\hat {\cal O}'_2\cdot E_\lambda^{(2)}(x_1^{1\over 2},x_2^{1\over 2})=\Lambda^{(2,2)}\cdot E_\lambda^{(2)}(x_1^{1\over 2},x_2^{1\over 2})
\ee
with the anzatz
\be
E_\lambda^{(2)}(x_1,x_2)=x^\lambda+\sum_{\mu<\lambda}C_{\lambda\mu}^{(2)}x^\mu
\ee
and $|\lambda|$ odd to realize that solutions are also numbered by \wcs! It completes the construction of 2-twisted non-symmetric Macdonald polynomials $E_\lambda^{(2)}(x_1,x_2)$.

\subsection{Properties of $E^{(2)}_\lambda$ at $n=2$}

Hence, we constructed another series of non-symmetric polynomials. Moreover, one can check that one again can obtain symmetric polynomials associated with the dominant integral weights from these non-symmetric polynomials by summing up over the Weyl group $W={\cal S}_n$, i.e. over all permutations of the partition $\lambda^+$:
\be
M^{(2)}_{\lambda^+}=\sum_{{\lambda=w\cdot\lambda^+}\atop{w\in W}} E^{(2)}_\lambda\cdot\left(\prod_{(i,j):\ \lambda_j>\lambda_i}{1-q^{\lambda_j-\lambda_i}t^{\zeta(\lambda)_i-\zeta(\lambda)_j-1}\over
1-q^{\lambda_j-\lambda_i}t^{\zeta(\lambda)_i-\zeta(\lambda)_j}}\right)
\ee
The product in the summand runs over pairs of $(i,j)$ such that $\lambda_i<\lambda_j$. This gives the symmetric 2-twisted Macdonald polynomials. Surprisingly, the coefficients in this formula coincide with those in (\ref{nss}). The explanation of this fact is the universality that we discuss below.

In fact, one can find a manifest formula for the polynomials in this case:

\noindent
at $\lambda_1\le \lambda_2$
\be
E_{[\lambda_1,\lambda_2]}^{(2)}(x_1,x_2)=(x_1x_2)^{\lambda_1}E_{[0,\lambda_2-\lambda_1]}^{(2)}(x_1,x_2)
\ee
and, at $\lambda_1\ge \lambda_2$,
\be
E_{[\lambda_1,\lambda_2]}^{(2)}(x_1,x_2)=(x_1x_2)^{\lambda_2}E_{[\lambda_1-\lambda_2,0]}^{(2)}(x_1,x_2)
\ee
with
\be
E_\lambda^{(2)}(x_1^{1\over 2},x_2^{1\over 2})=\Omega^{-1}\sum_{k=0}^\lambda
(-1)^kt^{-{k\over 2}-\lambda+1}q^{{(1-\lambda)k\over 2}}C_k^{(\lambda)}(\lambda;q,t)\Big({tx_1\over x_2};q\Big)_k
x_2^{{\lambda+k\over 2}}\Omega(q,tq^k;x_1,x_2q^{\lambda-k})
\ee
at $\lambda$ even and
\be
E_\lambda^{(2)}(x_1^{1\over 2},x_2^{1\over 2})=\bar\Omega^{-1}\sum_{k=0}^\lambda
(-1)^kt^{-{k\over 2}-\lambda+1}q^{{(1-\lambda)k\over 2}}C_k^{(\lambda)}(\lambda;q,t)\Big({tx_1\over x_2};q\Big)_k
x_2^{{\lambda+k\over 2}}\Omega(q,tq^k;x_1,x_2q^{\lambda-k})
\ee
at $\lambda$ odd. The coefficients $C_k$ are equal to
\be
C_k^{([0,\lambda])}(\lambda;q,t)={t(t-1)\over tq^k-1}\binom{\lambda-1}{k}_q
\prod_{i=1}^{\lambda-k-1}{t^{2}q^{\lambda-i}-1\over tq^{\lambda-i}-1}&\ \ \ \ \ &\hbox{for }E_{[0,\lambda]}^{(2)}\nn\\
C_k^{([\lambda,0])}(\lambda;q,t)=q^k \binom{\lambda}{k}_q\prod_{i=0}^{\lambda-k-1}{t^{2}q^{\lambda-i}-1\over tq^{\lambda-i}-1}&\ \ \ \ \ &\hbox{for }E_{[\lambda,0]}^{(2)}
\ee

These formulas have to be compared with the standard formula for the (untwisted) non-symmetric Macdonald polynomials \cite{Mac}:

\noindent
at $\lambda_1\le \lambda_2$
\be
E_{[\lambda_1,\lambda_2]}=(x_1x_2)^{\lambda_1}E_{[0,\lambda_2-\lambda_1]}
\ee
and, at $\lambda_1\ge \lambda_2$,
\be
E_{[\lambda_1,\lambda_2]}=(x_1x_2)^{\lambda_2}E_{[\lambda_1-\lambda_2,0]}
\ee
with
\be\label{E0a}
E_{[0,\lambda]}=\sum_{k=0}^{\lambda-1}x_1^kx_2^{\lambda-k}\binom{\lambda-1}{k}_q\ \prod_{i=1}^k{tq^{i-1}-1\over tq^{\lambda-i}-1}
\ee
and
\be\label{Ea0}
E_{[\lambda,0]}=\sum_{k=0}^\lambda x_1^{k}x_2^{\lambda-k}q^{\lambda-k}\binom{\lambda}{k}_q\ \prod_{i=0}^k{tq^{i}-1\over tq^{\lambda-i}-1}
\ee

\section{Eigenfunctions of twisted Cherednik Hamiltonians\label{efCH}}

Unfortunately, the things look so simple only in the lowest non-trivial case of $n=2$ and $a=2$,
though even in this case the factors $\Omega$ are different for even and odd levels.
For higher values of $n$ or $a$ interpretation in terms of Vandermonde-like twisting
does not persist. In this section, we construct eigenfunction of the twisted Cherednik Hamiltonians at arbitrary $a$. We mostly concentrated on the very explicit example of the two-particle system, and describe $n>2$ case in less detail.

\subsection{General solution at $n=2$}

At $n=2$ and higher $a$, the whole construction becomes different, and reduces to the construction of $a=2$ case described in the previous section in a peculiar way. The general polynomial solution of the eigenfunction equations at $t=q^{-m}$ is again based on the lowest grade solution $\Omega^{(a)}(m;x_1,x_2)$, which is
\be\label{EBA3}
\boxed{
\Omega^{(a)}(m;x_1,x_2)= \Psi_m^{(a)}\Big({a-2\over 2}m,{a\over 2}m;x_1,x_2\Big)
}=
\ee
\be
&=&\sum_{j=0}^{(a-1)m} q^{j^2-(a-1)mj\over a}  x_1^{j\over a}x_2^{(a-1)m-j\over a}
\sum_{k=0}^{[\frac{j}{a}]}(-1)^k
q^{(a-1)mk-(j-1)k+\frac{k(k-1)}{2}}\times\nn\\
&\times&
\frac{[m+j-ak-1]_q!}{  [j-ak]_q![m-k]_q![k]_q!}
 \left([m-k]_q + q^{j+m-(a+1)k} [k]_q\right)\nn
\ee
so that formula (\ref{EBA}) still persists. Again, how to continue this formula to arbitrary $t$ is not that clear, we know only that $\Omega^{(1)}(m;x_1,x_2)=1$, and $\Omega^{(2)}(m;x_1,x_2)$ is given by formula (\ref{Omegat}).

However, we assume that it can be continued, and note that all solutions that become at $t=q^{-m}$ polynomials of $x_{1,2}^{1\over a}$ are of the form:
\be\label{Psip}\boxed{
\psi_\lambda^{(a)}(q,t;x_1,x_2)=\sum_{k=0}^\lambda c_k(\lambda;q,t)\Big({tx_1\over x_2};q\Big)_kx_2^k\cdot\left[
t^{k\over a}(q^{-k}x_2)^{\lambda-k\over a}\Omega^{(a)}(q,tq^k;x_1,x_2q^{\lambda-k})\right]
}
\ee
at $\lambda<m$.

In particular, $\psi_0^{(a)}(q,t;x_1,x_2)=\Omega^{(a)}(q,t;x_1,x_2)$. Note that {\bf the dependence on $a$ is hidden only in the quantities $(q^{-k}x_2)^{{\lambda-k\over a}}t^{k\over a}\Omega^{(a)}(q,tq^k;x_1,x_2q^{\lambda-k})$}, which are effectively functions of $x_{1,2}^{1\over a}$ (polynomials of $x_{1,2}^{1\over a}$ at $t=q^{-m}$ and integer $m$): the coefficients $c_k(\lambda;q,t)$ do not depend on $a$, neither do the Pochhammer symbols.

In accordance with (\ref{fac}), one can definitely always multiply (\ref{Psip}) by the prefactor $(x_1x_2)^\alpha$, and it is still a solution. At a given $\lambda$, there are always only two ``basic'' solutions of the eigenvalue equations, which do not have a prefactor $(x_1x_2)^\alpha$:
\be\label{c12}
c^{(1)}_k&=&(-1)^kt^{-k-\lambda}q^{k(3-k)\over 2} \binom{\lambda}{k}_q\prod_{i=0}^{\lambda-k-1}{t^{2}q^{\lambda-i}-1\over tq^{\lambda-i}-1}\nn\\
c^{(2)}_k&=&
(-1)^kt^{-k-\lambda+1}q^{k(1-k)\over 2}{t-1\over tq^k-1}\binom{\lambda-1}{k}_q
\prod_{i=1}^{\lambda-k-1}{t^{2}q^{\lambda-i}-1\over tq^{\lambda-i}-1}
\ee
reducing at $t=q^{-m}$ to
\be
c^{(1)}_k&=&(-1)^kq^{{k(1-k)\over 2}+2mk} {\displaystyle{\binom{2m-k-1}{\lambda-k}_q}\over\displaystyle{\binom{m-k-1}{\lambda-k}_q}}\ \binom{\lambda}{k}_q\nn\\
c^{(2)}_k&=&(-1)^kq^{{k(1-k)\over 2}+2mk}{[m]\over [m-k]}
{\displaystyle{\binom{2m-k-1}{\lambda-k-1}_q}\over\displaystyle{\binom{m-k-1}{\lambda-k-1}_q}}\ \binom{\lambda-1}{k}_q
\ee
At $\lambda\ge m$ some of these coefficients become singular. Non-singular solutions in such cases, which can be obtained by a regularization do not have form (\ref{Psip}). Hence, at integer $m$ and $\lambda\ge m$ some of solutions disappear, but there are some other sporadic solutions\footnote{For instance, at $a=3$, $\lambda=1$, $m=1$, there is only one solution $x_2\Omega^{(3)}(q,q^{k-1};x_1,qx_2)$, and at $a=3$, $\lambda=2$, $m=1$, there is a ``sporadic" solution
$$
\psi_2^{(3)}(q,q^{-1};x_1,x_2)=q^{4\over 3}x_1^{1\over 3}x_2+q^{4\over 3}x_2^{4\over 3}-q^{1\over 3}x_1+q^{1\over 3}x_1x_2^{1\over 3}+2q^{1\over 3}x_1^{1\over 3}x_2+2x_1^{2\over 3}x_2^{2\over 3}
$$}
.

Of the two basic solutions (\ref{c12}), the second solution provides $c_\lambda^{(2)}=0$, i.e. it is proportional to $x_2^{1\over p}$. These two solutions are associated at $a=1,2$ with polynomials $E_{[\lambda,0]}$ and $E_{[0,\lambda]}$ correspondingly, and the normalization of $c_k$ is chosen in such a way that the corresponding polynomials have the proper normalization. For an illustration, we explain how this works in the case of $a=1$.

\subsection{Specialization to $a=1$}

Since $\Omega^{(1)}(q,t;x_1,x_2)=1$, one obtains from (\ref{Psip}) and (\ref{c12}) two eigenfunctions:
\be
\psi_\lambda=\sum_{k=0}^\lambda(-1)^kt^{-\lambda}q^{-k\lambda+{k(k+1)\over 2}} \binom{\lambda}{k}_q\Big({tx_1\over x_2};q\Big)_k
\left(\prod_{i=0}^{\lambda-k-1}{t^{2}q^{\lambda-i}-1\over tq^{\lambda-i}-1}\right)x_2^\lambda\stackrel{(\ref{Poch})}{=}\nn\\
=\sum_{k,j=0}^\lambda(-1)^{k+j}t^{j-\lambda}q^{-k\lambda+{k(k+1)\over 2}+{j(j-1)\over 2}} \binom{\lambda}{k}_q\binom{k}{j}_q
\left(\prod_{i=0}^{\lambda-k-1}{t^{2}q^{\lambda-i}-1\over tq^{\lambda-i}-1}\right)x_1^jx_2^{\lambda-j}
\ee
and
\be
\bar\psi_\lambda=\sum_{k=0}^{\lambda-1}(-1)^kt^{-\lambda+1}q^{-k(\lambda-1)+{k(k+1)\over 2}}{t-1\over tq^k-1}\binom{\lambda-1}{k}_q\Big({tx_1\over x_2};q\Big)_k
\left(\prod_{i=1}^{\lambda-k-1}{t^{2}q^{\lambda-i}-1\over tq^{\lambda-i}-1}\right)x_2^\lambda\stackrel{(\ref{Poch})}{=}\nn\\
=\sum_{k,j=0}^{\lambda-1}(-1)^{k+j}t^{j-\lambda+1}q^{-k(\lambda-1)+{k(k+1)\over 2}+{j(j-1)\over 2}}{t-1\over tq^k-1}\binom{\lambda-1}{k}_q\binom{k}{j}_q
\left(\prod_{i=1}^{\lambda-k-1}{t^{2}q^{\lambda-i}-1\over tq^{\lambda-i}-1}\right)x_1^jx_2^{\lambda-j}
\ee
where the sums over $j$ run up to $j=k$.

These formulas can be simplified: the double sums can be reduced to single sums using the two related identities
\be
\sum_{k=j}^\lambda(-1)^kq^{{k(k+1)\over 2}-k\lambda}
\binom{\lambda}{k}_q\binom{k}{j}_q\prod_{i=0}^{\lambda-k-1}{t^{2}q^{\lambda-i}-1\over tq^{\lambda-i}-1}=(-1)^j(qt)^{\lambda-j}q^{-{j(j-1)\over 2}}\binom{\lambda}{j}_q\prod_{i=0}^j{tq^i-1\over tq^{\lambda-i}-1}\nn\\
\sum_{k=j}^\lambda(-1)^kq^{{k(k+1)\over 2}-k\lambda}{t-1\over tq^k-1}
\binom{\lambda}{k}_q\binom{k}{j}_q\prod_{i=0}^{\lambda-k-1}{t^{2}q^{\lambda-i}-1\over tq^{\lambda-i}-1}=(-1)^jt^{\lambda-j}q^{-{j(j-1)\over 2}}\binom{\lambda}{j}_q\prod_{i=0}^{j-1}{tq^i-1\over tq^{\lambda-i}-1}
\ee

Finally, the result reads
\be
\psi_\lambda=\sum_{k=0}^\lambda x_1^{k}x_2^{\lambda-k}q^{\lambda-k}\binom{\lambda}{k}_q\ \prod_{i=0}^k{tq^{i}-1\over tq^{\lambda-i}-1}\stackrel{(\ref{Ea0})}{=}E_{[\lambda,0]}\nn\\
\bar\psi_\lambda=\sum_{k=0}^{\lambda-1}x_1^kx_2^{\lambda-k}\binom{\lambda-1}{k}_q\ \prod_{i=1}^k{tq^{i-1}-1\over tq^{\lambda-i}-1}\stackrel{(\ref{E0a})}{=}E_{[0,\lambda]}
\ee

\subsection{Extension to higher $n$}

The general polynomial solution of the eigenfunction equations at $t=q^{-m}$ is again based on the lowest grade solution\footnote{We remind that one can freely multiply this solution by an arbitrary power $\alpha$ of $x_1x_2\ldots x_n$, which results into the shift
$$
(x_1x_2\ldots x_n)^\alpha\Psi_m^{(a)}\Big(0,m,\ldots,(n-1)m;x_1,\ldots,x_n\Big)= \Psi_m^{(a)}\Big(\alpha,m+\alpha,\ldots,(n-1)m+\alpha;x_1,\ldots,x_n\Big)
$$
and the new function is still a solution.}
$\Omega^{(a)}(m;x_1,\ldots,x_n)$
\be\label{EBAn}
\boxed{
\Omega^{(a)}(m;x_1,\ldots,x_n)\sim\Psi_m^{(a)}\Big(0,m,\ldots,(n-1)m;x_1,\ldots,x_n\Big)
}
\ee
However, solutions are now looking a bit more tricky.

Consider, for instance, the case of $n=3$.

At level $|\lambda|=1$, there are three eigenfunctions associated with three possible weak compositions [0,0,1], [0,1,0] and [1,0,0]. They have the form
\be\label{n3}
\psi_{[0,0,1]}^{(a)}&=&x_3^{1\over a}\Omega^{(a)}(q,t;x_1,x_2,qx_3)\nn\\
\psi_{[0,1,0]}^{(a)}&=&{\Big\{{tx_2\over x_3}\Big\}\over \Big\{{x_2\over x_3}\Big\}}
x_2^{1\over a}\Omega^{(a)}(q,t;x_1,qx_2,x_3)+
{(1-t)\over 1-qt^2}{\Big\{{qt^2x_3\over x_2}\Big\}\over \Big\{{x_3\over x_2}\Big\}}x_3^{1\over a}\Omega^{(a)}(q,t;x_1,x_2qx_3)\nn
\\
\psi_{[1,0,0]}^{(a)}&=&{\Big\{{tx_1\over x_2}\Big\}\Big\{{tx_1\over x_3}\Big\}\over \Big\{{x_1\over x_2}\Big\}\Big\{{x_1\over x_3}\Big\}}x_1^{1\over a}\Omega^{(a)}(q,t;qx_1,x_2,x_3)+
{(1-t)\over (1-qt)}{\Big\{{tx_2\over x_3}\Big\}\Big\{{qtx_2\over x_1}\Big\}\over \Big\{{x_2\over x_3}\Big\}\Big\{{x_2\over x_1}\Big\}}x_2^{1\over a}\Omega^{(a)}(q,t;x_1,qx_2,x_3)+\nn\\
&+&{(1-t)\over (1-qt)}{\Big\{{tx_3\over x_2}\Big\}\Big\{{qtx_3\over x_1}\Big\}\over \Big\{{x_3\over x_1}\Big\}\Big\{{x_3\over x_2}\Big\}}x_3^{1\over a}\Omega^{(a)}(q,t;x_1,x_2,qx_3)
\ee
where we use the notation $\{x\}:=1-x$.
Note that {\bf these solutions are still polynomials at $t=q^{-m}$}, though it is not evident at all: this is a peculiar property of the Baker-Akhiezer function $\Omega^{(a)}$, which deserves further studying. One can see that the coefficients in each of these expressions contain the same number of fractions, and this number in $\psi_{\lambda}^{(a)}$ is determined by the minimal number of permutations needed to obtain $\lambda$ from $w_0\lambda$, where $w_0$ is the longest permutation in permutation group ${\cal S}_n$. For instance, in $\psi_{[0,0,1]}^{(a)}$ there are no fractions, since $[0,0,1]=w_0[1,0,0]$. In $\psi_{[0,1,0]}^{(a)}$ there is one fraction, and [0,1,0] is obtained from [0,0,1] by one permutation minimally, while in $\psi_{[1,0,0]}^{(a)}$ there are two fractions, and [1,0,0] is obtained from [0,0,1] minimally by two permutations.

Similarly, at level $|\lambda|=2$, the simplest eigenfunctions are
\be
\psi_{[0,1,1]}^{(a)}&=&(x_2x_3)^{1\over a}\Omega^{(a)}(q,t;x_1,qx_2,qx_3)\\
\psi_{[1,0,1]}^{(a)}&=&{(1-t)\over (1-qt^2)}{\Big\{{qt^2x_2\over x_1}\Big\}\over \Big\{{x_2\over x_1}\Big\}}(x_2x_3)^{1\over a}\Omega^{(a)}(q,t;x_1,qx_2,qx_3)+
{\Big\{{tx_1\over x_2}\Big\}\over \Big\{{x_1\over x_2}\Big\}}(x_1x_3)^{1\over a}\Omega^{(a)}(q,t;qx_1,x_2,qx_3)\nn
\ee
Introduce the quantity
\be\label{Xi}
\Xi^{(a)}_\lambda:=\left(\prod_{i=1}x_i^{\lambda_i\over a}q^{\lambda_i(\lambda_i-1)\over 2a}\right)\ \Omega^{(a)}(q,t;\{q^{\lambda_i}x_i\})=
\prod_{i=1}\left(x_i^{1\over a}q^{\hat D_i}\right)^{\lambda_i}\ \Omega^{(a)}(q,t;\{x_i\})
\ee
Thus defined quantity automatically takes into account that multiplying a solution with a factor of
$\prod_{i=1}^nx_i^\alpha$ gives rise to another solution, and shifts all entries in the weak composition by $\alpha$: $\lambda_i\to\lambda_i+\alpha$. This follows from the property $\Omega^{(a)}(\{q^\alpha x_i\})\sim
\Omega^{(a)}(\{x_i\})$. Thus, now one can immediately deal with all possible weak compositions, not obligatory with those having at least one zero part.

In this notation, for instance,
\be
\psi_{[1,1,0]}^{(a)}&=&{\Big\{{tx_1\over x_3}\Big\}\over \Big\{{x_1\over x_3}\Big\}}
{\Big\{{tx_2\over x_3}\Big\}\over \Big\{{x_2\over x_3}\Big\}}\ \Xi^{(a)}_{[1,1,0]}+
{(1-t)\over (1-qt)}\left(
{\Big\{{tx_1\over x_2}\Big\}\over \Big\{{x_1\over x_2}\Big\}}{\Big\{{qtx_3\over x_2}\Big\}\over \Big\{{x_3\over x_2}\Big\}}\
\Xi^{(a)}_{[1,0,1]}+{\Big\{{tx_2\over x_1}\Big\}\over \Big\{{x_2\over x_1}\Big\}}
{\Big\{{qtx_3\over x_1}\Big\}\over \Big\{{x_3\over x_1}\Big\}}\ \Xi^{(a)}_{[0,1,1]}\right)\\
\psi_{[0,0,2]}^{(a)}&=&{\Big\{{qtx_3\over x_2}\Big\}\over \Big\{{qx_3\over x_2}\Big\}}{\Big\{{qtx_3\over x_1}\Big\}\over\Big\{{qx_3\over x_1}\Big\}}\Xi^{(a)}_{[0,0,2]}+{(1-t)\over (1-qt)}
\left({\Big\{{tx_1\over x_3}\Big\}\over \Big\{{x_1\over qx_3}\Big\}}{\Big\{{tx_1\over x_2}\Big\}\over\Big\{{x_1\over x_2}\Big\}}\Xi^{(a)}_{[1,0,1]}+{\Big\{{tx_2\over x_3}\Big\}\over \Big\{{x_2\over qx_3}\Big\}}{\Big\{{tx_2\over x_1}\Big\}\over\Big\{{x_2\over x_1}\Big\}}\Xi^{(a)}_{[0,1,1]}\right)\nn
\ee
From these examples one could expect that the coefficient in front of any $\Xi^{(a)}_\mu$ is always a product of a few fractions.
However, the next eigenfunction demonstrates that this is not the case: one of the coefficients in $\psi_{[0,2,0]}^{(a)}$ (that in front of $\Xi^{(a)}_{[0,1,1]}$) becomes a sum of two terms, each of them still being a product of three fractions:
\be\label{020}
\psi_{[0,2,0]}^{(a)}&=&{\Big\{{qtx_2\over x_3}\Big\}\over \Big\{{qx_2\over x_3}\Big\}}{\Big\{{qtx_2\over x_1}\Big\}\over\Big\{{qx_2\over x_1}\Big\}}{\Big\{{tx_2\over x_3}\Big\}\over\Big\{{x_2\over x_3}\Big\}}\Xi^{(a)}_{[0,2,0]}+{(1-t)\over (1-q^2t^2)}
{\Big\{{q^2t^2x_3\over x_2}\Big\}\over \Big\{{qx_3\over x_2}\Big\}}{\Big\{{qtx_3\over x_2}\Big\}\over\Big\{{x_3\over x_2}\Big\}}{\Big\{{qtx_3\over x_1}\Big\}\over\Big\{{qx_3\over x_1}\Big\}}\Xi^{(a)}_{[0,0,2]}+\nn\\
&+&{q(1-t)\over (1-qt)(1+t)}
\left({(1+q)(1-qt^2)\over(1-q^2t^2)}
{\Big\{{qt^2x_3\over x_2}\Big\}\over \Big\{{qx_3\over x_2}\Big\}}{\Big\{{tx_2\over x_3}\Big\}\over\Big\{{qx_2\over x_3}\Big\}}
{\Big\{{tx_2\over x_1}\Big\}\over\Big\{{x_2\over x_1}\Big\}}+
{\Big\{{tx_2\over x_3}\Big\}\over \Big\{{qx_2\over x_3}\Big\}}{\Big\{{t^2x_3\over x_1}\Big\}\over\Big\{{x_3\over x_1}\Big\}}
{\Big\{{tx_1\over x_2}\Big\}\over\Big\{{x_1\over x_2}\Big\}}\right)
\Xi^{(a)}_{[0,1,1]}+\nn\\
&+&{(1-t)^2\over (1-qt)(1-q^2t^2)}{\Big\{{q^2t^2x_3\over x_2}\Big\}\over \Big\{{x_3\over x_2}\Big\}}{\Big\{{tx_1\over x_3}\Big\}\over\Big\{{x_1\over qx_3}\Big\}}{\Big\{{tx_1\over x_2}\Big\}\over \Big\{{x_1\over x_2}\Big\}}\Xi^{(a)}_{[1,0,1]}+
{(1-t)\over(1-qt)}
{\Big\{{tx_2\over x_3}\Big\}\over \Big\{{x_2\over x_3}\Big\}}{\Big\{{tx_1\over x_3}\Big\}\over\Big\{{x_1\over x_3}\Big\}}{\Big\{{tx_1\over x_2}\Big\}\over \Big\{{x_1\over qx_2}\Big\}}\Xi^{(a)}_{[1,1,0]}
\ee
At last, the sixth remaining eigenfunction at this level is
\be\label{200}
\psi_{[2,0,0]}^{(a)}&=&{\Big\{{qtx_1\over x_3}\Big\}\over \Big\{{qx_1\over x_3}\Big\}}{\Big\{{qtx_1\over x_2}\Big\}\over\Big\{{qx_1\over x_2}\Big\}}{\Big\{{tx_1\over x_2}\Big\}\over \Big\{{x_1\over x_2}\Big\}}{\Big\{{tx_1\over x_3}\Big\}\over \Big\{{x_1\over x_3}\Big\}}\Xi^{(a)}_{[2,0,0]}+{(1-t)\over(1-q^2t)}\left(
{\Big\{{tx_2\over x_3}\Big\}\over \Big\{{x_2\over x_3}\Big\}}{\Big\{{qtx_2\over x_3}\Big\}\over\Big\{{qx_2\over x_3}\Big\}}{\Big\{{qtx_2\over x_1}\Big\}\over \Big\{{qx_2\over x_1}\Big\}}{\Big\{{q^2tx_2\over x_1}\Big\}\over \Big\{{x_2\over x_1}\Big\}}\Xi^{(a)}_{[0,2,0]}+\right.\nn\\
&+&\left.{\Big\{{tx_3\over x_2}\Big\}\over \Big\{{x_3\over x_2}\Big\}}{\Big\{{qtx_3\over x_2}\Big\}\over\Big\{{qx_3\over x_2}\Big\}}{\Big\{{qtx_3\over x_1}\Big\}\over \Big\{{qx_3\over x_1}\Big\}}{\Big\{{q^2tx_3\over x_1}\Big\}\over \Big\{{x_3\over x_1}\Big\}}\Xi^{(a)}_{[0,0,2]}\right)+{q(1+q)(1-t)^2\over(1-qt)(1-q^2t)}
{\Big\{{tx_3\over x_2}\Big\}\over \Big\{{qx_3\over x_2}\Big\}}{\Big\{{tx_2\over x_3}\Big\}\over\Big\{{qx_2\over x_3}\Big\}}{\Big\{{qtx_3\over x_1}\Big\}\over \Big\{{x_3\over x_1}\Big\}}{\Big\{{qtx_2\over x_1}\Big\}\over \Big\{{x_2\over x_1}\Big\}}\Xi^{(a)}_{[0,1,1]}+\nn\\
&+&{q(1+q)(1-t)\over (1-q^2t)}
\left({\Big\{{tx_3\over x_2}\Big\}\over \Big\{{x_3\over x_2}\Big\}}{\Big\{{tx_1\over x_3}\Big\}\over\Big\{{qx_1\over x_3}\Big\}}{\Big\{{qtx_3\over x_1}\Big\}\over \Big\{{qx_3\over x_1}\Big\}}{\Big\{{tx_1\over x_2}\Big\}\over \Big\{{x_1\over x_2}\Big\}}\Xi^{(a)}_{[1,0,1]}+
{\Big\{{tx_2\over x_3}\Big\}\over \Big\{{x_2\over x_3}\Big\}}{\Big\{{tx_1\over x_2}\Big\}\over\Big\{{qx_1\over x_2}\Big\}}{\Big\{{qtx_2\over x_1}\Big\}\over \Big\{{qx_2\over x_1}\Big\}}{\Big\{{tx_1\over x_3}\Big\}\over \Big\{{x_1\over x_3}\Big\}}\Xi^{(a)}_{[1,1,0]}\right)
\ee
Note that the pattern with the number of fraction in each term persists: in $\psi_{[0,1,1]}^{(a)}$ there are no fractions, in $\psi_{[1,0,1]}^{(a)}$ there is one ([1,0,1] is obtained from [0,0,1] minimally by one permutation), in $\psi_{[1,1,0]}^{(a)}$ there are two fractions ([1,1,0] is obtained from [0,0,1] minimally by two permutation). Similarly, as soon as in $\psi_{[0,0,2]}^{(a)}$ there are two fractions, in $\psi_{[0,2,0]}^{(a)}$ there are three, and in $\psi_{[2,0,0]}^{(a)}$ there are four fractions.

The same structure of eigenfunctions emerges at higher $n$ and $\lambda$. Thus, one can naturally expect the general formula for the eigenfunction to be of the form (notice the triangular structure)
\be\label{mainn}
\boxed{
\psi_\lambda=\sum_{\mu\le\lambda}F_{\lambda\mu}(x)\ \Xi^{(a)}_\mu
}
\ee
and {\bf all $F_{\lambda\mu}(x)$'s do not depend on $a$.} 

Each $F_{\lambda\mu}(x)$ is a homogeneous rational function of $x_i$'s, which generally is a sum of products of the form
\be\label{rat}
F_{\lambda\mu}(x)\sim\sum\prod_{(i,j)}^{N_\lambda} {\Big\{{a_{ij}tx_i\over x_j}\Big\}\over \Big\{{b_{ij}x_i\over x_j}\Big\}}
\ee
and the number of fractions $N_{\lambda}$ in these products is the same for all $\mu$ at fixed $\lambda$, and $N_{\lambda^+}-N_{\lambda}$ is equal to the minimal length of permutation that brings the weak composition $\lambda$ to $\lambda^+$. Here $a_{ij}$, $b_{ij}$ are monomials of $q$ and $t$. Moreover, {\bf at ${\bf t=q^{-m}}$}, the coefficients in front of products (\ref{rat}) entering $F_{\lambda\mu}(x)$ are ratios of $q$-numbers\footnote{From this, it is clear from the very beginning that $F_{[0,0,2][0,1,1]}(x)$ in (\ref{020}) contains two terms: the coefficient in front of $x_2x_3$ in (\ref{Ens}) is not a ratio of $q$-numbers because of the factor $(1+qt-qt^2-q^2t^2)$, which is a sum of two $q$-numbers, and implies a non-trivial multiplicity in this case.} (up to possible monomials of $q$), and {\bf $\psi_\lambda$ is a polynomial (!),} while individual terms in the sum at each $\mu$ are not.

Denote the number of solutions of the eigenvalue equations through $p(\lambda,n)$. We checked at $t=q^{-m}$ and at various values of $a$, $n$, $\lambda$ and $m>|\lambda|$ that $p(\lambda,n)$ does not depend on $a$ and $m$.
\
Taking into account this fact and presumable formula (\ref{mainn}),

\bigskip

\framebox{\parbox{15cm}{
{\bf We conjecture that, at generic $n$, all eigenfunctions are given by formula (\ref{mainn}) with the twist $a$ entering the formula only through the functions $\Xi_\lambda^{(a)}$ in (\ref{Xi}), i.e. through the functions $\Omega^{(a)}$ shifted and multiplied by proper monomials made of $x_i^{1\over a}$.}}}

\bigskip

We tested this conjecture at various particular values of $a$ and $n$, it perfectly works.

\subsection{Properties of eigenfunctions}

Earlier, we listed typical properties of sets of non-symmetric Macdonald polynomials. They are basically the same as in the symmetric polynomial case: stability, triangular structure, orthogonality, Cauchy identity. There is also formula (\ref{nss}) that makes symmetric polynomials. All these properties are expected to preserve for the eigenfunctions due to the proposed universality!

In particular, {\bf the stability property}, i.e. the reduction
\be
\psi_{[\lambda_1,\ldots,\lambda_{n-1},\lambda_n]}^{(a)}(x_1,\ldots,x_{n-1},0)=\left\{
\begin{array}{cr}\psi_{[\lambda_1,\ldots,\lambda_{n-1}]}^{(a)}(x_1,\ldots,x_{n-1})&\ \ \ \ \ \lambda_n=0\cr
0&\ \ \ \ \ \lambda_n\ne 0
\end{array}
\right.
\ee
follows from the property of the Baker-Akhiezer functions (at $t=q^{-m})$):
\be
\Psi_m^{(a)}\Big(0,m,\ldots,(n-2)m,(n-1)m;x_1,\ldots,x_{n-1},0\Big)=
\left(\prod_{i=1}^{n-1}x_i\right)^{m(a-1)\over a}\Psi_m^{(a)}\Big(0,m,\ldots,(n-2)m;x_1,\ldots,x_{n-1}\Big)\nn
\ee
i.e.
\be
\Omega^{(a)}(x_1,\ldots,x_{n-1},0)=
\left(\prod_{i=1}^{n-1}x_i\right)^{m(a-1)\over a}\Omega^{(a)}(x_1,\ldots,x_{n-1})
\ee
and from the triangular structure (\ref{mainn}).

In its turn, {\bf the triangular structure} is a direct corollary of the universality, while {\bf the orthogonality} is induced by the orthogonality of the Baker-Akhiezer functions (the CMM formulas \cite{ChE}), and {\bf the Cauchy identity} follows from the orthogonality.

At last, the counterpart of formula (\ref{nss}) giving rise to {\bf symmetric functions}
associated with the dominant integral weights from the non-symmetric eigenfunctions and obtained by summing up over the Weyl group $W={\cal S}_n$, i.e. over all permutations of the partition $\lambda^+$ is expected to be
\be
{\cal M}^{(a)}_{\lambda^+}=\sum_{{\lambda=w\cdot\lambda^+}\atop{w\in W}} \psi^{(a)}_\lambda\cdot\left(\prod_{(i,j):\ \lambda_j>\lambda_i}{1-q^{\lambda_j-\lambda_i}t^{\zeta(\lambda)_i-\zeta(\lambda)_j-1}\over
1-q^{\lambda_j-\lambda_i}t^{\zeta(\lambda)_i-\zeta(\lambda)_j}}\right)
\ee
The product in the summand runs over pairs of $(i,j)$ such that $\lambda_i<\lambda_j$. ${\cal M}^{(a)}_{\lambda^+}$ becomes at $t=q^{-m}$ a symmetric polynomial of $x_i^{1\over a}$.
We checked this formula in simple examples, it works, and this is quite natural because of the universality.

Note, however, that these symmetric functions are not eigenfunctions of the DIM Hamiltonians, since the eigenvalues corresponding to eigenfunctions associated with distinct weak compositions of the same $\lambda^+$ are distinct.

\subsection{Eigenvalues}

In the case of $n=2$, the vector of eigenvalues
from \eqref{Cev} is naturally parameterized by the two numbers
$\lambda_1$ and $\lambda_2$.
\be \label{eq:n-2-arb-a-eigens}
  \Lambda_{\lambda_1,\lambda_2}^{(a,\cdot)} = (q^{\mu_1}, q^{\mu_2});
  \ \ (\mu_1, \mu_2) =
  \left(\frac{2(a-1)a + 1}{4 a}, \frac{2(a-1)a + 1}{4 a}\right) + (\lambda_1, \lambda_2)
  + \left\{\begin{array}{l}
    \lambda_1 < \lambda_2: \ \ (a m, (a-1)m)
    \cr
    \lambda_1 \geq \lambda_2: \ \ ((a-1)m, a m)
  \end{array} \right.
\ee
Extension to the generic $n$ is immediate. For instance, at $n=3$,
\begin{align} \label{eq:n-3-arb-a-eigens}
  \Lambda_{\lambda_1,\lambda_2,\lambda_3}^{(a,\cdot)} = & \ (q^{\mu_1}, q^{\mu_2},q^{\mu_3});
  \ \ (\mu_1, \mu_2,\mu_3) =
  \left(\frac{2(a-1)a + 1}{4 a}, \frac{2(a-1)a + 1}{4 a},\frac{2(a-1)a + 1}{4 a}\right)
  + (\lambda_1,\lambda_2,\lambda_3)
  \\ \notag
  & \quad\quad\quad\quad\quad\quad\quad\quad\quad\quad\quad\quad
  + \sigma_{\lambda}((2 (a-1) + 0) m, (2 (a-1) + 1) m, (2 (a-1) + 2) m),
\end{align}
where $\sigma_\lambda$ is the \textit{minimal} permutation that brings $\lambda$
to $\lambda^+$.

\bigskip

Note that these eigenvalues can be obtained even before evaluating the eigenfunctions from the Jack limit, see the Appendix.

Note also that for some $m$ one or more degenerations occur: several sets $\lambda$ turn out
to have the same vector of eigenvalues. Then we superficially have multidimensional
solution spaces, where it is not straightforward to find distinguished basis.
These ambiguities are, however, resolved for high enough $m$.

\section{Conclusion}

\subsection{Summary
\label{summ}}

In this paper, we described an extension of the toy but basic example of sec.\ref{toy}, applied to standard systems of non-symmetric polynomials, to systems of eigenvalues associated with the $N$-body representation of DIM algebra and with the twisted Cherednik algebra. That is,

\begin{itemize}

\item{}  The starting point is the choice of commuting operators
$\hat c_i$, $i = 1, \ldots N$ and their power sums $\hat  h_k$.
In DIM/Cherednik case, these are {\bf commuting} $a$-twisted Cherednik operators $\mathfrak{C}_i^{(a)}:=\hat C_i x_i \hat C_i x_i \ldots x_i \hat C_i$, which are
the product of $a$ ``rotated" Cherednik operators (\ref{calCi}), i.e. those having the grading -1, and $a-1$ variables $x_i$
having grading $1$. The power sums of the $a$-twisted Cherednik operators, when acting on symmetric functions are equal to the DIM algebra Hamiltonians $\hat H^{(a)}_k$ associated with the integer rays $(-1,a)$ \cite{MMP}.

\item{}  Since the grading of these operators is $-1$, they do not have polynomial eigenfunctions.
This is, however, compensated by conjugation with $q^{\frac{1}{2a}\sum_{j=1}^N z_j^2}$.

\item{}
These rotated eigenfunctions are polynomials only at $t=q^{-m}$ with $m\in\mathbb{N}$.

\item The are polynomials of the fractional powers of the variables, $x_i^{1/a}$.

\item Among the eigenfunctions, there is a kind of ``ground state" $\Omega^{(a)}$: the eigenfunction with minimal grading. At $t=q^{-m}$ with $m\in\mathbb{N}$, this eigenfunction is a symmetric function of $x_i$ (as the ground state has to be) and, hence, it is simultaneously an eigenfunction of the DIM algebra Hamiltonians $\hat H^{(a)}_k$. Their eigenfunctions are the twisted Baker-Akhiezer functions \cite{ChE,ChF,MMP1}, which are generally not symmetric. However, $\Omega^{(a)}$ is proportional to the twisted Baker-Akhiezer functions at special values of parameters, when it is symmetric. The grading of $\Omega^{(a)}$ is equal to $m(a-1) \cdot \frac{n(n-1)}{2}$.

\item{}
The generic eigenfunctions are labeled by \wcs\ $\lambda$,
but their grading is now shifted from $|\lambda|$ to
$|\lambda| + m(a-1) \cdot \frac{n(n-1)}{2}$.

\item{}
For low values of $m\le |\lambda|$ some eigenfunctions merge, and one needs consideration at larger $m$. For instance, at $m=1$, $\psi_{[1,0,0]}^{(a)}$ coincides with $\psi_{[0,1,0]}^{(a)}$, see (\ref{n3}).

\item  {\bf The generic eigenfunctions can be realized (\ref{mainn}) as linear sums of $\Omega^{(a)}$
(multiplied by proper monomials made of $x_i^{1\over a}$ and $q^{1\over a}$) with expansion functions (rational functions of $x_i$) that do not depend on $a$. Hence, only the ground state functions $\Omega^{(a)}$ control a peculiar twisting. This universality reflects an $SL(2,\mathbb{Z})$ symmetry of the DIM (automorphism Miki \cite{Miki1}) and Cherednik algebras.}

\item{}
The pattern of eigenfunctions gets very explicit and transparent in the limit of $q\rightarrow 1$,
which is basically of the first order in $\hbar = \log q$. In this limit, the twisted Hamiltonians are reduced to untwisted ones by a conjugation with a simple Vandermonde-like function so that the eigenfunctions are just multiplied by this function.

\end{itemize}

\subsection{Discussion}

Section \ref{summ} formulates the conclusions, resulting from our difficult search for eigenfunctions
of the twisted Cherednik Hamiltonians.
The difficulty is not just technical, but rather conceptual.
The reason is that the answer lies beyond the comfortable
world of symmetric polynomials and essentially relies on non-symmetric ones.
The theory of these latter is vast (see an extensive list of references in \cite{Alex}), but it has yet nothing like the beauty and the power of the former.
In particular, no generalization of Fock representation exists, i.e. that in terms of power sums $p_k=\sum_ix_i^k$.

Now let us list the problems that have to be studied further.

\begin{itemize}

\item{}
One of the first things to do in the future is to start a physics-oriented description(s)
of the theory of non-symmetric polynomials.

\item The eigenfunctions of the twisted Cherednik system are conjectured to be described by formula (\ref{mainn}). However, the explicit form of the rational functions $F_{\lambda\mu}(x)$ yet to be further specified in order to achieve at arbitrary $n$ the concreteness similar to formula (\ref{Psip}) in the case of $n=2$.

\item The ground state solution of the system as we established is a peculiar Baker-Akhiezer function. It is a non-trivial property of this kind of Baker-Akhiezer function that formulas like (\ref{mainn}) becomes polynomial at $t=q^{-m}$. An origin of this very non-trivial property remains unclear.

\item{}
The next need is description of rational rays $(b,a)$ of \cite{MMP} with coprime $a$ and $b>1$.
This seems relatively straightforward, still we avoid too far-going and not-well-enough-grounded speculations,
before the problem is studied in more detail.

\item{}
All this implies certain rethinking of integrability theory, where both $n$-particle quantum mechanics
and eigenvalue matrix models are no longer providing the fully adequate interpretations, since, in most studied examples,
they are both restricted to the sets of symmetric polynomials, in particle coordinates and eigenvalues respectively.
The first attempts of such generalizations appeared in the form of {\it triad} in \cite{triad} (see also further extensions in \cite{MMPZ1,MMPZ2}),
relating standard symmetric eigenfunctions to non-symmetric Baker-Akhiezer functions and non-polynomial (and non-symmetric)
Noumi-Shiraishi power series.
The present paper gives a much broader and, in a sense, a more fundamental view on the situation.
Still, we are just at the beginning of this new non-symmetric journey into
the (super)integrability ($\stackrel{?}{=}$ non-perturbative physics) world.

\end{itemize}

\noindent
After this comprehensive introduction and unification of three subjects:
integrability theory inspired by the DIM algebras, integrability theory inspired by the twisted Cherednik algebras, and {\it non-symmetric} polynomials, we look for forthcoming achievements in this promising field.
There are plenty of smaller problems which need to be addressed and resolved.

\section*{Acknowledgements}

This work is supported by the RSF grant 24-12-00178.

\bigskip

\bigskip

\section*{Appendix.
Limit of $q\to 1$ at twisted Cherednik systems
}

In this Appendix, we discuss the limit of $q\to 1$, $t=q^\beta$ keeping $\beta$ fixed, which gives rise to a kind of twisted non-symmetric Jack polynomials. However, as we shall explain, they differ from the non-twisted ones only by a simple factors. Still, looking at them allows one to understand the structure of the twisted non-symmetric Macdonald polynomials better.

Throughout the Appendix, {\bf we choose $\beta=-m$ with $m\in\mathbb{Z}_{\ge 0}$, and use the notation $X_i:=x_i^{1\over a}$.}

\subsection*{Limit of Cherednik operators} \label{sec:poisson-limit}

The limit of $q=1$ is actually about the first order in $\hbar:=\log q$, because the zeroth-order
is not sensitive to eigenfunctions: action on any function would be just unity.
Thus, the actual limit is not quite trivial.
Technically it uses the following definitions instead of the first lines of (\ref{ordefs})
\be
r_{ij} = 1 - am\hbar\cdot\frac{X_j^a}{X_i^a-X_j^a}(1-\sigma_{ij}), \ \ \ \ \
r^{-1}_{ij} = 1 + amh\cdot \frac{X_j^a}{X_i^a-X_j^a}(1-\sigma_{ij}), \ \ \ \ \
q^{\hat D_i} = 1 + \hbar X_i\frac{\p}{\p X_i}
\ee
and the Cherednik operators in the lower lines of (\ref{ordefs}) are calculated up to the first order in $\hbar$.

Let us study the limit of eigenvalue problem \eqref{Cev}. That is, consider the eigenvalue
problem for the operator limit
\newcommand\mD[0]{\mathfrak{D}}
\begin{align} \label{eq:poisson-limit-def}
  \mD^{(a)}_i f(\ux) := \lim_{\hbar\rightarrow0}
  \frac{\br{q^{-\frac{1}{2 a} \sum_i z_i^2}
      \mfC^{(a)}_i
      q^{\frac{1}{2 a} \sum_i z_i^2}
    \Bigg{|}_{\substack{q=e^\hbar \\ t=e^{-m\hbar}}}
    - 1}}{\hbar}  f(\ux)
  = \mu^{(a,i)} f(\ux)
\end{align}

\bigskip

For the sake of simplicity, we consider the case of $n=2$. The limit operators $\mD_1$ and $\mD_2$
are manifestly equal to\footnote{Note that for all ``twists'' $a$ such defined operators $\mD_i$
are first order differential operators, and therefore, for $a > 2$, cannot be associated with
the Yangian counterparts of $\mfC^{(a)}_i$ (see \cite[Eq.(34)]{MMCal}, \cite[Eq.(79)]{MMMP1}) that are $a$-th order differential operators.
The question of how to take the DIM $\longrightarrow$ Yangian limit in this setup,
as well as the question about eigenfunctions for the twisted Dunkl operators themselves
are very intriguing and deserve a separate study.
}
\begin{align}
  \mD_1 = a x_1 \frac{\partial}{\partial x_1}
  + \frac{\br{(a-1)^2 + a^2}}{4 a} \cdot I
  + \frac{m x_2^{\frac{1}{a}}}{(x_1^{\frac{1}{a}}-x_2^{\frac{1}{a}})} \cdot \sigma_{1,2}
  - \frac{x_2}{(x_1 - x_2)} a m \cdot I
  \\ \notag
  \mD_2 = a x_1 \frac{\partial}{\partial x_1}
  + \frac{\br{(a-1)^2 + a^2}}{4 a} \cdot I
  - \frac{m x_2^{\frac{1}{a}}}{(x_1^{\frac{1}{a}}-x_2^{\frac{1}{a}})} \sigma_{1,2}
  + \frac{x_1}{(x_1 - x_2)} a m \cdot I
\end{align}

\bigskip

As before, we search for simultaneous eigenfunctions for both $\mD_1$ and $\mD_2$
as homogeneous polynomials of some degree $d$ in the ``fractional'' variables
$X_1=x_1^{\frac{1}{a}}, X_2=x_2^{\frac{1}{a}}$.
at degrees $0 .. (a-1) m - 1$ there are no solutions, and the unique solution
at degree $(a-1)m$ is given by
\begin{align}
  \Omega_0^{(a)} (X_1,X_2) = \frac{\br{X_1^a - X_2^a}^m}{\br{X_1 - X_2}^m}
\end{align}
Solutions at higher degrees are all \textit{proportional} to $\Omega_0^{(a)} (X_1,X_2)$,
so it makes sense to consider conjugated operators
\begin{align}
  \widetilde{\mD}_1 = \Omega_0^{-1} \mD_1 \Omega_0
  = & \
  X_1 \frac{\partial}{\partial X_1}
  + \br{
    \underbrace{\frac{\br{(a-1)^2 + a^2}}{4 a} + (a-1)m}
    _{\text{can be shifted away}} - m\frac{X_2}{\br{X_1 - X_2}}}\cdot I
  + \frac{m X_2}{\br{X_1-X_2}} \cdot \sigma_{1,2}
  \\ \notag
  \widetilde{\mD}_2 = \Omega_0^{-1} \mD_2 \Omega_0
  = & \
  X_2 \frac{\partial}{\partial X_2}
  + \br{
    \underbrace{\frac{\br{(a-1)^2 + a^2}}{4 a} + (a-1)m}
    _{\text{can be shifted away}} + m\frac{X_1}{\br{X_1 - X_2}}}\cdot I
  - \frac{m X_2}{\br{X_1-X_2}} \cdot \sigma_{1,2}
\end{align}
where we express differentiation in terms of $X_i$ as well.
After the trivial shift
\begin{align}
  \widetilde{\mD}_i \rightarrow \widetilde{\mD}_i -
  \br{\frac{\br{(a-1)^2 + a^2}}{4 a} + (a-1)m} \cdot I
\end{align}
the operators $\widetilde{\mD}$
\textbf{no longer depend on $a$} and, in fact, are equal to the
Dunkl limit of the (vertical) Cherednik operators (\ref{Dl}):
\begin{align}
  D_i := \lim_{\hbar\rightarrow0}
  \frac{\br{C_i \Bigg{|}_{\substack{q=e^\hbar \\ t=e^{\beta\hbar}}}
    - 1}}{\hbar}
  =
  \left\{
  \begin{array}{l}
    i=1: \ \ x_1 \frac{\partial}{\partial x_1} + \beta \frac{x_2}{\br{x_1-x_2}}
    \br{I - \sigma_{1,2}}
    \\
    i=2: \ \ x_2 \frac{\partial}{\partial x_2} - \beta \frac{x_1}{\br{x_1-x_2}} \cdot I
    + \beta \frac{x_2}{\br{x_1-x_2}} \cdot \sigma_{1,2}
  \end{array}
  \right.
\end{align}
provided $\beta = -m$. Therefore their eigenfunctions (for all $a$) are nothing but the non-symmetric Jack polynomials.

\bigskip

We, therefore, conclude that the limit of $q\to 1$ \eqref{eq:poisson-limit-def}
of the eigenvalue problem for the twisted Cherednik operators turns out to be much simpler
than the full problem:
all dependence on $a$ is contained in the common factor $\Omega_0$,
the shift of eigenvalues, and the change of variables $x_i\to X_i = x_i^{\frac{1}{a}}$. This phenomenon is, in fact, known for the limit of $q\to 1$ of the twisted Baker-Akhiezer functions \cite{Ano}.

\subsection*{Limit eigenfunctions}

\bigskip

As we observed the eigenfunctions in the limit of $q\to 1$ are not much different from the non-symmetric Jack polynomials. The structure of eigenfunctions in this limit system is

\begin{itemize}

\item{}
At level $|\lambda|=0$, the eigenfunction is
\be
\Omega_0^{(a)} =  \left(\prod_{i<j}^n \frac{X_i^a - X_j^a}{X_i-X_j}\right)^m
\label{VDMatQ1}
\ee
with the eigenvalue
\be
\mu_{\Omega_0^{(a)}}^{(i)}=q^{m\big((a-1)n+i-a\big)+\frac{a-1}{2}}
\ee

\item{}
The simplest eigenfunction at level $|\lambda|=1$ is
\be
\psi_{[0,\ldots,0,1]}\cdot \Delta = X_n\cdot \Delta
\ee
with the eigenvalues
\be
\mu^{(a,i)}=\mu_{\Omega_0^{(a)}}^{(i)} \cdot q^{-(nm-1)\delta_{i,n}}
\ee
Naturally, the eigenvalue for the $n$-th operator (with $i=n$) in this case differs from the others.

\item{}
The total number of eigenfunctions at level $|\lambda|=1$ is $n$,
up to possible coincidences of distinct eigenfunctions at particular values of parameters $n,a,m$.
These are:
\be
\psi_{0,\ldots,0,1}:&=&X_n, \nn \\
\psi_{0,\ldots,0,1,0}:&=& \Big((n-1)m-1\Big)X_{n-1}+mX_n, \nn \\
\psi_{0,\ldots,0,1,0,0}:&=& \Big((n-2)m-1\Big)X_{n-2}+mX_{n-1}+mX_n, \nn \\
\psi_{0,\ldots,0,1,0,0,0}:&=& \Big((n-3)m-1\Big)X_{n-3}+mX_{n-2}+mX_{n-1}+mX_n, \nn \\
\psi_{0,\ldots,0,1,0,0,0,0}:&=& \Big((n-4)m-1\Big)X_{n-4}+mX_{n-3}+mX_{n-2}+mX_{n-1}+mX_n, \nn \\
&\ldots& \nn \\
\psi_{0,0,1,0,\ldots,0}:&=& (3m-1)X_3+mX_4+\ldots +mX_n \nn \\
\psi_{0,1,0,\ldots,0}:&=& (2m-1)X_2+mX_3+mX_4+\ldots +mX_n \nn \\
\psi_{1,0,\ldots,0}:&=& (m-1)X_1+mX_2+mX_3+mX_4+ \ldots + mX_n
\ee
As one can see, there are no degenerations at level $|\lambda|=1$ except for the case of $m=1$,
when the last function in the list,  $\psi_{1,0,\ldots,0} = (m-1)X_1+mX_2+mX_3\ldots + mX_n$,
becomes independent of $X_1$ and coincides with next to the last one,
$\psi_{0,1,0,\ldots,0}= (2m-1)X_2+mX_3+\ldots +mX_n$.

For
\be
\psi_{0,\ldots,0,\underbrace{1}_s,0, \ldots, 0}
= (sm-1)X_s+m\sum_{s'=s+1}^n X_{s'}
\ee
the  $i$-th eigenvalue is
\be
\mu^{(a,i)}_\lambda=\mu_{\Omega_0^{(a)}}^{(i)}\cdot q^{-(sm-1)\delta_{i,s}+m\cdot {\bf he}(s-i)}
\ee
where the Heaviside function ${\bf he}(x)=1$ for $x\geq 0$ and ${\bf he}(x)=0$ for $x<0$
(thus $i=s$ is present in the both terms in the exponent).

\item{}
At level $|\lambda|=2$, examples are provided by
\be
\psi_{0,\ldots, 0,1,1} \sim &X_{n-1}X_n, \nn \\
\psi_{0,\ldots,0,2,0 } \sim &m  \Big(\big(m(n-1)-2\big)X_{n-1} + m X_n\Big)\sum_{i=1}^{n-2} X_i
+\nn \\
&  + (m-1) \Big(\big(m(n-1)-2\big)X^2_{n-1} + m X^2_n\Big) + m\big(mn-2\big)X_{n-1}X_n
\label{exalevel2}
\ee
Two more, $\psi_{2,0,\ldots, 0}$ and $\psi_{0,\ldots,0,2}$ can be extracted by substitution of $|\lambda|=2$
from general formulas (\ref{psiL0}) and (\ref{psi0L}) below.
Still these are only four out of the $\frac{n(n+1)}{2}$ eigenfunctions at this level.

For generic $n$,  $\psi_{0,\ldots, 0,1,1} = X_{n-1}X_n$ is always an eigenfunction
with the eigenvalue ???
$$\mu^{(a,i)}=\mu_{\Omega_0^{(a)}}^{(i)}\cdot q^{2m\sum_{j=1} \delta_{i,j}{\bf he}(n/2-j)
- \big((n-2)m-1\big)\sum_j^n\delta_{i,j}{\bf he}(j-n/2)}???$$
Actually the first sum runs over $j=1,\ldots,entier\left(\frac{n+1}{2}\right)$.

\item{} At $n=2$, one can write down explicit formulas for the eigenvalues. This is in no way a surprise, since explicit formulas for the non-symmetric Jack polynomials in this case are immediately obtained from (\ref{E0a}) and (\ref{Ea0}) in the $q\to 1$ limit. The answers are
\be
\psi^{(2,a,m)}_{L,0} \sim  \sum_{j=0}^L  \frac{L!}{j!(L-j)!}\frac{m!(m-L-1)!}{(m-j-1)!(m+j-L)!}  X_1^{j}X_2^{L-j}
\label{psi2amL0}
\ee
and
\be
\psi^{(2,a,m)}_{0,L} \sim
X_2 \sum_{j=0}^{L-1}  \frac{(L-1)!}{j!(L-1-j)!}\frac{m!(m-L)!}{(m-j)!(m+j-L)!}  X_1^{j}X_2^{L-1-j}
\ee
Note that the second function is proportional to $X_2$.
For $L\geq m$, these formulas can look singular, but actually they are not, as can be seen
by expressing factorials through $\Gamma$-functions.
Numerators in these formulas do not affect $X$-dependence, and, in this sense, are irrelevant.

A link between (\ref{psi2amL0}) and the limit of (\ref{c12}) is provided by a peculiar identity
\be
\!\!\!\!\! \!\!\!\!\!\!\!\!\!\!\! \left(\frac{X_1^a-X_2^a}{X_1-X_2}\right)^m \!\!
 \left\{\sum_{j=0}^|\lambda|
 \frac{|\lambda|!}{j!(|\lambda|-j)!}\frac{(m-|\lambda|-1)!}{(m-j-1)!} X_2^{|\lambda|-j}
 \left( \frac{m!}{(m+j-|\lambda|)!}X_1^j  - \frac{(2m-j-1)!}{(2m-|\lambda|-1)!} \,(X_1-X_2)^j\right)
 \right\}
 = 0\nn
\ee

\item{}
For generic $n$, one gets instead of (\ref{psi2amL0})
\be
\psi^{(n,a,m)}_{[|\lambda|,0,\ldots,0]} \sim  \sum_{k,j_2,\ldots,j_n=0}^|\lambda|
\frac{|\lambda|! \ \delta_{k+j_2+\ldots+j_n,|\lambda|}}{k!j_2!\ldots j_n!}
\frac{X_1^{k}\prod_{s=2}^n X_s^{j_s}}{(m-1-k)!\prod_{s=2}^n (m-j_s)!}
\ee
Note that the $m$-dependent factor is not invariant under the permutations of  $k=j_1$ and all other $j_s$.
This is the basic origin of asymmetry of the polynomial
(despite, in this case, the \wc\ $[|\lambda|,0,\ldots,0]$ is actually a Young diagram),
which will only increase for other excitations.

Another way to write the same formula is (up to total normalization)
\be
\boxed{
\psi^{(n,a,m)}_{[|\lambda|,0,\ldots,0]} \sim  \sum_{k,j_2,\ldots,j_n=0}^|\lambda|  \delta_{k+j_2+\ldots+j_n,|\lambda|}\cdot   \frac{ X_1^{k}}{k!(m-1-k)!}\prod_{s=2}^n \frac{X_s^{j_s}}{j_s! (m-j_s)!}
 \ \ \ \ {\rm with\ e.v.}\
\mu_{\Omega_0^{(a)}}^{(i)}\cdot q^{|\lambda|\delta_{i,1}}
}
\label{psiL0}
\ee

\item{}
However, generic excitations for $n>2$ is now much trickier.
First, we need expressions for the other single-column \wcs \ $[0,\ldots,0,|\lambda|,0, \ldots 0]$.
Second, we need expressions for all \wcs\ which have vanishing entries.
And only those with all non-vanishing entries will be reduced by separation of the factors $\prod_{i=1}^n X_i$.

In fact, in the $q\to 1$ limit, these are not too complicated expressions,
for example,
\be
\psi^{(n,a,m)}_{[0,\ldots,0,|\lambda|]} \sim   \sum_{k,j_2,\ldots,j_n=0}^{|\lambda|-1}  \delta_{k+j_2+\ldots+j_n,|\lambda|-1}\cdot   \frac{ X_1^{k}}{k!(m-k)!}\left(\prod_{s=2}^{n-1} \frac{X_s^{j_s}}{j_s! (m-j_s)!})\right)   \frac{X_n^{j_n+1}}{j_n! (m-j_n-1)!}
\label{psi0L}
\ee
It is proportional to $X_n$ and has eigenvalues $\mu^{(a,i)}=\mu_{\Omega_0^{(a)}}^{(i)}\cdot q^{m}\cdot q^{(|\lambda|-mn)\delta_{i,n}}$.
There are additional simplifications, well illustrated by the example (\ref{exalevel2}).
Like there, all $\psi_{\underbrace{0,\ldots,0}_k,\underbrace{1,\ldots,1}_{n-k}} = \prod_{i=k+1}^n X_i$.
Still the majority of eigenfunctions are not so simple.
\end{itemize}

To summarize, the pattern of eigenfunctions in the limit of $q\to 1$ is very simple and clear.

\begin{itemize}

\item{}
They turn to be nicely separated from the background Vandermonde-like factor $\Omega_0$ (\ref{VDMatQ1}), i.e.
all  eigenfunctions look like $\psi_\lambda = \Omega_0^{(a)} \cdot J_\lambda$
and eigenvalues are $\mu^{(a,i)}_\lambda=\mu_{\Omega_0^{(a)}}^{(i)} \mu_{J_\lambda}^{(i)}$.

\item{}
$J$ are just the non-symmetric Jack polynomials
independent of the twisting parameter $a$,
while $a$-dependence persists in $\Omega_0^{(a)}$.

\item{}
In the limit of $q\rightarrow 1$, the eigenvalues arise in the form $1+\hbar\xi$,
but can be easily continued to $\mu=q^\xi$, where they coincide with the
true eigenvalues for an arbitrary $q$.
Such continuation does not hold for the twisted Cherednik eigenfunctions themselves,
which are in general neither factorizable,
nor $a$-independent.

\item{}
Still the number of eigenfunctions, as well as degeneration rules for particular $m$,
when some $\psi$ coincide, are fully seen in the limit of $q\rightarrow 1$.

\item Beyond the limit of $q\to 1$, the naive factorization $\psi_\lambda = \Omega_0^{(a)} \cdot J_\lambda$ fails,
and one can need a more sophisticated twisting, as we demonstrated in the main body of the paper.

\end{itemize}


\bigskip

\bigskip

\begin{thebibliography}{12}

\bibitem{Jack}  H. Jack, Proc. R. Soc. Edinburgh (A) {\bf 69} (1970) 1; (1972) 347

\bibitem{St} R.P. Stanley, Adv. Math. {\bf 77} (1988) 76

\bibitem{Turb} W. Ruehl and A. Turbiner, Mod. Phys. Lett. {\bf A10} (1995) 2213, hep-th/9506105

\bibitem{Mac} I.G. Macdonald,
  \textit{Symmetric functions and Hall polynomials},
  Oxford University Press, 1995

\bibitem{UFN2} A. Morozov,
Phys.Usp.(UFN) {\bf 37} (1994) 1,  hep-th/9303139;
hep-th/9502091; hep-th/0502010

\bibitem{UFN3} A. Mironov, Int.J.Mod.Phys. {\bf A9} (1994) 4355; Phys.Part.Nucl.
{\bf 33} (2002) 537, hep-th/9409190;
Electron. Res. Announ. AMS \textbf{9} (1996) 219-238,
hep-th/9409190

\bibitem{DI} J. Ding, K. Iohara, 
Lett. Math. Phys. {\bf 41} (1997) 181-193, q-alg/9608002

\bibitem{Miki} K. Miki, J. Math. Phys. {\bf 48} (2007) 123520

\bibitem{K} M. Kapranov,  
Algebraic geometry {\bf 7}, J.Math. Sci. {\bf 84} (1997) 1311-1360, alg-geom/9604018

\bibitem{BS} I. Burban, O. Schiffmann, 
Duke Math. J. {\bf 161} (2012) 1171, arXiv:math/0505148

\bibitem{S}  O. Schiffmann, 
J. Algebraic Combin. {\bf 35} (2012) 237-26, arXiv:1004.2575

\bibitem{Feigin} B. Feigin, M. Jimbo, T. Miwa, E. Mukhin,
Commun. Math. Phys. \textbf{356}  (2017) 285, arXiv:1603.02765

\bibitem{MMP} A. Mironov, A. Morozov, A. Popolitov, JHEP \textbf{09} (2024) 200,
  arXiv:2406.16688

\bibitem{triad} A.~Mironov, A.~Morozov, A.~Popolitov,
Phys. Lett. B \textbf{869} (2025) 139840,
arXiv:2411.16517

\bibitem{Cha} O. Chalykh,  Adv.Math. {\bf 166(2)} (2002) 193-259, math/0212313

\bibitem{ChE}   O. Chalykh, P. Etingof, Advances in Mathematics, {\bf 238} (2013) 246-289,
arXiv:1111.0515

\bibitem{MMP1} A.~Mironov, A.~Morozov, A.~Popolitov,
Phys. Lett. B \textbf{863} (2025) 139380,
arXiv:2410.10685

\bibitem{NS} M. Noumi, J. Shiraishi, 
arXiv:1206.5364

\bibitem{DIMDAHA} P. Di Francesco, R. Kedem, Comm. Math. Phys. {\bf 369(3)} (2019) 867-928, arXiv:1208.4333

\bibitem{Ch} I. Cherednik,
  Vol. {\bf 319}. Cambridge University Press, 2005.

\bibitem{NSCh}
  M. Nazarov,~E. Sklyanin,
  IMRN, vol.2019, iss.8 (2019) pp.2266–2294; arXiv:1703.02794

\bibitem{ChF}
O. Chalykh, M. Fairon, 
J.Geom.Phys. {\bf 121} (2017) 413-437,
  arXiv:1704.05814

\bibitem{Alex} \url{https://www.symmetricfunctions.com/}

\bibitem{HHL} J. Haglund, M. Haiman, N. Loehr, American Journal of Mathematics, {\bf 130(2)} (2008) 359-383, math/0601693

\bibitem{KN} An. N. Kirillov, M. Noumi, Duke Math. J. {\bf 93(1)} (1998) 1-39, q-alg/9605004

\bibitem{MMkn} A.~Mironov, A.~Morozov,
arXiv:2508.07255

\bibitem{CO} K. Mimachi, M. Noumi, 
Duke Math. J. {\bf 95(1)} (1998) 621-634, q-alg/9610014

\bibitem{Las03} A. Lascoux, {\sl Double crystal graphs}, In: Studies in Memory of Issai Schur, Birkhäuser Boston, 2003

\bibitem{Zenk} Y.~Zenkevich,
JHEP \textbf{03} (2023) 193,
arXiv:2112.14687

\bibitem{Miki1} K. Miki, 
Lett. Math. Phys. {\bf 47} (1999) 365-378

\bibitem{Zenk2} Y.~Zenkevich,
JHEP \textbf{08} (2021) 149,
arXiv:1812.11961

\bibitem{MMPf} A.~Mironov, A.~Morozov, A.~Popolitov,
Nucl. Phys. B \textbf{1012} (2025) 116809,
arXiv:2411.14194

\bibitem{MMPdet} A.~Mironov, A.~Morozov, A.~Popolitov,
Eur. Phys. J. C \textbf{85} (2025) no.5, 574,
arXiv:2504.02022

\bibitem{MMPZ1} A.~Mironov, A.~Morozov, A.~Popolitov, Z.~Zakirova,
Phys. Lett. B \textbf{865} (2025) 139467,
arXiv:2412.19588

\bibitem{MMPZ2} A.~Mironov, A.~Morozov, A.~Popolitov, Z.~Zakirova,
Pisma Zh. Eksp. Teor. Fiz. \textbf{121} (2025) no.9, 788-795,
arXiv:2503.07592

\bibitem{MMCal} A.~Mironov, A.~Morozov,
Phys. Lett. \textbf{B842} (2023) 137964,
arXiv:2303.05273

\bibitem{MMMP1}  A.~Mironov, V.~Mishnyakov, A.~Morozov, A.~Popolitov,
JHEP \textbf{23} (2020) 065,
arXiv:2306.06623

\bibitem{Ano} A. Anokhina, to appear

\end{thebibliography}
\end{document}